\documentclass[12pt]{article}
\usepackage[reqno]{amstex}
\usepackage{amscd}
\newcommand{\beq}{\begin{equation}}
\newcommand{\eeq}{\end{equation}}
\newcommand{\RG}{{\mathcal R}}
\newcommand{\RGB}{{\mathcal R}^{\mathrm Bol}}
\newcommand{\RGV}{{\mathcal R}^{\mathrm Pot}}
\newcommand{\SG}{{\mathcal S}}
\newcommand{\wt}[1]{\widetilde{#1}}
\newcommand{\dm}[1]{{\rm d}\mu_{\Gamma_L}(#1)}
\newcommand{\pl}{L\frac{\partial}{\partial L}}
\newcommand{\pdx}{x\frac{\partial}{\partial x}}
\newcommand{\pdy}[1]{x_{#1}\frac{\partial}{\partial x_{#1}}}
\newcommand{\pdp}{p\frac{\partial}{\partial p}}
\newcommand{\pdq}[1]{p_{#1}\frac{\partial}{\partial p_{#1}}}
\newcommand{\ppl}{\frac{\delta}{\delta\psi_L}}
\newcommand{\pp}{\frac{\delta}{\delta\psi}}
\newcommand{\pd}{\frac{\delta}{\delta\phi}}
\newcommand{\DG}{{\mathcal D}}
\newcommand{\CG}{{\mathrm C}}
\newcommand{\R}{{\mathbb R}}
\newcommand{\N}{{\mathbb N}}
\newcommand{\Z}{{\mathbb Z}}
\newcommand{\ZZ}{{\mathcal Z}}
\newcommand{\VV}{{\mathcal V}}
\newcommand{\KK}{{\mathcal K}}
\newcommand{\OBS}{{\mathcal O}}
\newcommand{\ddx}[1]{{\rm d}^Dx_1\dots{\rm d}^Dx_{#1}}
\newcommand{\ddp}[1]{\frac{{\rm d}^Dp_1}{(2\pi )^D}\dots
 \frac{{\rm d}^Dp_{#1}}{(2\pi)^D}}
\newcommand{\px}[1]{\phi(x_1)\dots\phi(x_{#1})}
\newcommand{\LUV}{{\Lambda_{{\rm UV}}}}
\newcommand{\LIR}{{\Lambda_{{\rm IR}}}}
\begin{document}
\begin{titlepage}
\begin{center}
{\Large The renormalized $\phi^4_4$-trajectory by \\[2mm]
perturbation theory in a running coupling \\[2mm]
using partial differential equations} \\[10mm]
\end{center}
\begin{center}
{\Large C. Wieczerkowski} \\[10mm]
\end{center}
\begin{center}
Institut f\"ur Theoretische Physik I,
Universit\"at M\"unster, \\
Wilhelm-Klemm-Stra\ss e 9, D-48149 M\"unster, \\
wieczer$@$yukawa.uni-muenster.de \\[2mm]
\end{center}
\vspace{-10cm}
\hfill MS-TP1-96-03
\vspace{11cm}
\begin{abstract}
We compute the renormalized trajectory of $\phi^4_4$-theory
by perturbation theory in a running coupling. We use an
exact infinitesimal renormalization group. The expansion is
put into a form which is manifestly independent of the
scale parameter.
\end{abstract}
\end{titlepage}

\section{Introduction}

In Wilson's renormalization group \cite{W71,WK74}, renormalized
theories come as renormalized trajectories of effective actions.
The renormalization group leaves invariant a renormalized action
up to a flow of renormalization parameters. This transformation
law {\it defines} a renormalized theory. It opens a way to a
formulate a renormalized theory without the detour to a limit
procedure.

We study the $\phi^4$-theory in four dimensional Euclidean
space time. A dimension parameter $D$ will be kept in the
equations to display the dimension dependence of scale factors.
We use a renormalization group transformation $\RG_L$ which
depends on a scale parameter $L$. The scale parameter is
equal to the ratio of an ultraviolet and an infrared cutoff.
We use an exact renormalization group differential equation
for the $L$-dependence of effective actions. On the renormalized
$\phi^4$-trajectory the effective interaction depends on $L$
through an $L$-dependent running coupling $g(L)$ only. A
natural choice for $g$ is the effective $\phi^4$-coupling.
On the renormalized trajectory, the normal ordered effective
potential $\VV (\phi,g)$ (see (\ref{norpot})) satisfies
the renormalization group differential equation
\begin{equation}
\left\{ \beta (g)\frac{\partial}{\partial g}-
\left(\DG\phi,\pd\right)\right\} \VV (\phi,g)=
-\left\langle \VV (\phi,g), \VV (\phi,g)\right\rangle.
\label{basic}
\end{equation}
It is the main dynamical equation in our approach. Its most
important property is {\it independence} of $L$. The
renormalization group leaves invariant the renormalized
trajectory up to a flow of $g$. This flow is encoded in a
$\beta$-function through an ordinary differential equation
\begin{equation}
L\frac{{\rm d}}{{\rm d}L}g(L)=\beta (g(L)).
\label{couflo}
\end{equation}
The operator $\left(\DG\phi,\pd\right)$ generates a scale
transformation. The right hand side of (\ref{basic}) is a
bilinear renormalization group bracket $\left\langle A(\phi),
B(\phi)\right\rangle$ defined in (\ref{biliterm}). It
consists of contractions between $A(\phi)$ and $B(\phi)$,
and is independent of $g$. The $\phi^4$-trajectory is the
solution to (\ref{basic}) with
\begin{equation}
\VV (\phi,g)= g \frac1{4!} \int{\rm d}^Dx \phi(x)^4+
g \frac{\zeta^{(1)}}{2} \int{\rm d}^Dx \partial_\mu \phi(x)
\partial_\mu \phi(x)+O(g^2).
\label{init}
\end{equation}
It is unique in a space of {\it finite} solutions. To be
precise, it is unique to all orders of perturbation theory
in $g$. Solution is here meant as a formal power series
in $g$ with polynomial coefficients in $\phi$. The
criterion of finiteness is that the polynomials are given
by smooth kernels on momentum space. Precise definitions
are given in the bulk of this paper. The first order wave
function term in (\ref{init}) is peculiar to the four
dimensional case. The coupling $\zeta^{(1)}$ follows when
(\ref{basic}) is expanded to second order in $g$.

The program of this paper is an iterative solution of
(\ref{basic}) and of (\ref{couflo}) in powers of $g$. In
its course we do not encounter any divergencies. The result
is a perturbation theory for the $\phi^4$-trajectory which
is finite to every order in $g$. This approach was proposed
in \cite{Wi96} using an iterated transformation with fixed
$L$. In this paper we investigate the dependence upon
smooth variation of $L$. It gives a refined formulation
based on the differential equation (\ref{basic}).
We restrict our attention to a one parametric family of
renormalized potentials. It is distinguished by the
property that no other vertices are included besides
those generated dynamically from the first order
$\phi^4$-vertex. The outcome is then a one-dimensional
curve, the $\phi^4$-trajectory. We write it as a curve
of effective potentials whose ultraviolet cutoff is
rescaled to unity. To make contact with the physical world
we would have to supply both a physical scale $\Lambda$
and a value of $g$.

We included elementary details to make this paper
self-contained. The experts in the field are begged pardon
and asked to jump directly to what we call local
perturbation theory. This paper is organized as follows.
In section one we formulate the setup, make the connection
with \cite{Wi96}, and give a derivation of the
infinitesimal renormalization group. Two comments on the
ultraviolet limit in the rescaled setup and the
relation between Green's functions and the effective
interaction are included. In section two we translate the
renormalization group flow into the normal ordered
representation. The second section closes with a bound on
the bilinear renormalization group bracket. In section
three we perform first an un-renormalized global
perturbation expansion as a comparison, and thereafter
the renormalized cutoff-free local perturbation expansion.
Explicit second order calculations are included to
demonstrate the methods. In the section four, the
renormalization group equations are solved. We present a
weak bound on the large momentum growth of the Euclidean
Green's functions. It ensures that the iteration is
indeed free of divergencies.

\section{Renormalization Group}

The setup of this paper is a momentum space renormalization
group for a Euclidean scalar field $\phi$.

\subsection{Renormalization Group Transformation}

Consider the following renormalization group transformation
$\RG_L$, depending on a scale parameter $L>1$. Let $\RG_L$
be composed of a Gaussian fluctuation integral, with covariance
$\Gamma_L$ and mean $\psi$, and a dilatation $\SG_L$ of
$\psi$. Let the fluctuation covariance be defined by
\begin{equation}
\wt{\Gamma}_L(p)=\frac 1{p^2}
\left\{\wt{\chi}(p)-\wt{\chi}(Lp)\right\},
\label{flucov}
\end{equation}
where $\wt{\chi}(p)$ is a momentum space cutoff function. A
convenient choice is the exponential cutoff
\begin{equation}
\wt{\chi}(p)=\exp \left(-p^2\right).
\end{equation}
It will be used in the following. Other choices however work
as well, for instance Pauli-Villar-regularization. The cutoff
function's purpose is to suppress momenta outside a momentum
slice $L^{-1}\leq |p|\leq 1$. Eq. (\ref{flucov})
defines a parameter dependent positive operator $\Gamma_L$
on the subspace of $L_2(\R^D)$ consisting of functions $f(x)$
with zero mode $\wt{f}(0)=0$. Let $\dm{\zeta}$ be the
associated Gaussian measure with mean zero on field space.
Recall its basic property
\begin{equation}
\int\dm{\zeta}
\exp\left\{(\zeta,f)\right\}=
\exp\left\{ \frac 12(f,\Gamma_L f)\right\},
\label{gauss}
\end{equation}
and consult for example Glimm and Jaffe \cite{GJ87} for further
information.

Let the fluctuation integral of a Boltzmann factor
$Z(\phi)=\exp\{-V(\phi)\}$ be defined as the average with
respect to (\ref{gauss}), shifted by an external field
$\psi$. A convenient notation for this average is
\begin{equation}
\left\langle Z\right\rangle_{\Gamma_L,\psi}=
\int\dm{\zeta} Z(\psi+\zeta).
\label{fluint}
\end{equation}
This fluctuation integral can be derived from a multi-scale
decomposition of a free massless scalar field.
Multi-scale decompositions are reviewed by Gallavotti in
\cite{G85}. See also the recent lectures \cite{BG95} by
Benfatto and Gallavotti, and references therein. The momentum
slice $L^{-1}\leq |p|\leq 1$ can be thought to label a rescaled
portion of momentum space degrees of freedom. This portion is
integrated out in (\ref{fluint}). The integration of another
portion is prepared for by a dilatation $\SG_L$ of $\psi$. Let
the dilatation be given by
\begin{equation}
\SG_L\psi (x)=L^{1-\frac D2}\psi\left(\frac xL\right).
\label{dilpsi}
\end{equation}
The exponent $\sigma=1-D/2$ is the scaling dimension of a
free massless field. Anomalous rescaling will not be considered
below. Non-anomalous rescaling applies at least to weak
perturbations of a free field.

The renormalization group transformation is the composition of
(\ref{fluint}) with (\ref{dilpsi}). For the Boltzmann factor it
reads
\begin{equation}
\RGB_L Z(\psi)=
\left\langle Z\right\rangle_{\Gamma_L,\SG_L\psi}.
\label{rgb}
\end{equation}
It is not difficult find renormalization group flows where
the Boltzmann factor $Z(\phi)$ is a polynomial in the field.
In scalar field theory the matter of interest are non-polynomial
flows of the form $Z(\phi)=\exp \{-V(\phi)\}$, where the potential
$V(\phi)$ is approximately local and polynomial. In the following
it is approximately quartic. The renormalization group
transformation for the potential reads
\begin{equation}
\RGV_L V(\psi)=
-\log\left\langle\exp (-V)\right\rangle_{\Gamma_L,\SG\psi}.
\label{rgv}
\end{equation}
The superscripts ${}^{\mathrm Bol}$ and ${}^{\mathrm Pot}$
will be dropped in the following by neglect of notation.
The below analysis will be done entirely in terms of the
potential. The matter of stability bounds on the Boltzmann
factor will not be addressed. The method will be perturbation
theory. It is valid in some vicinity of the trivial fixed
point $V_{*}(\phi)=0$. The renormalization group transformation
for the Boltzmann factor is identical with the linearized
transformation for the potential at the trivial fixed point.
The linearized renormalization group is responsible for the
leading dynamical behavior at weak coupling. A number of
elementary properties of (\ref{rgv}) are conveniently derived
from (\ref{rgb}), which is why the Boltzmann factor is
introduced here at all.

We restrict our attention to even potentials with
$V(-\phi)=V(\phi)$. Notice that field parity is preserved by
(\ref{rgv}). Potentials differing by a field independent
constant will be identified. We could also impose a normalization
condition, for instance $V(0)=0$. To maintain normalization
(\ref{rgv}) then has to be supplemented with subtraction of
$\RG_L V(0)$. Technically this constant is proportional to
the volume, infinite in infinite volume. Therefore, (\ref{rgv})
requires an intermediate volume cutoff to make sense. We
will wipe this technicality under the carpet and leave
(\ref{rgv}) as it stands. This setup is identical with that
in \cite{Wi96} up to the scale parameter $L$, which is
here variable.

\subsection{Semi-Group Property}

The composition of two renormalization group transformations with
scale $L$ is equal to one renormalization group transformation
with scale $L^2$. Moreover, the renormalization group
transformation (\ref{rgv}) satisfies
\begin{gather}
\RG_{L_1} \RG_{L_2}=\RG_{L_1 L_2}, \quad L_1, L_2>1,
\label{group} \\
lim_{L\rightarrow 1^+} \RG_L = {\mathrm id}.
\label{ident}
\end{gather}
The renormalization group therefore defines a representation of
the semi-group of dilatations of $\R^D$ with scale factors
$L>1$ on the space of effective interactions. The proof of
this semi-group property is
\begin{eqnarray}
\RG_{L_1}\RG_{L_2}Z(\psi)&=&
\int{\rm d}\mu_{\Gamma_{L_1}}(\zeta_1)
\int{\rm d}\mu_{\Gamma_{L_2}}(\zeta_2)
\;Z\left(\SG_{L_2}(\SG_{L_1}\psi+\zeta_1)+\zeta_2\right)
\nonumber \\ &=&
\int{\rm d}\mu_{\SG_{L_2}\Gamma_{L_1}\SG_{L_2}^T}(\zeta_1)
\int{\rm d}\mu_{\Gamma_{L_2}}(\zeta_2)
\;Z\left(\SG_{L_2}\SG_{L_1}\psi+\zeta_1+\zeta_2\right)
\nonumber \\ &=&
\int{\rm d}\mu_{\SG_{L_2}\Gamma_{L_1}\SG_{L_2}^T+\Gamma_{L_2}}(\zeta)
\;Z\left(\SG_{L_2}\SG_{L_1}\psi+\zeta\right)
\nonumber \\ &=&
\RG_{L_1L_2}Z(\psi).
\end{eqnarray}
It uses a dilatation- and a convolution identity for Gaussian
measures. Both follow from the functional Fourier transform
(\ref{gauss}). It also uses the elementary properties
\begin{gather}
\SG_{L_2}\Gamma_{L_1}\SG_{L_2}^T+\Gamma_{L_2}=\Gamma_{L_1L_2}, \\
\SG_{L_1}\SG_{L_2}=\SG_{L_1L_2}
\end{gather}
of (\ref{flucov}) and (\ref{dilpsi}). The property (\ref{ident})
follows from $\lim_{L\rightarrow 1^+}\wt{\Gamma}_L(p)=0^+$,
turning (\ref{gauss}) into a functional $\delta$-measure.

Let us remark that $\wt{\Gamma}_L(p)$ becomes negative for $0<L<1$.
Gaussian integration with negative covariance requires further
regularity of functionals on field space. Non-invertibility of
the renormalization group has been recently emphasized by
Benfatto and Gallavotti in \cite{BG95}. Effective theories tend to
require less parameters because the number of effective degrees of
freedom shrinks in the course of a renormalization group flow.
Inverse renormalization group transformations can however be given
a meaning on the renormalized trajectory. It is one dimensional
from the beginning. Furthermore, an interaction on the renormalized
trajectory is always the renormalization group image of another
one. Restricted to the renormalized trajectory, the renormalization
group defines a representation of the full dilatation group.

Due to the semi-group property the iteration of renormalization
group transformations with fixed scale is identical with an increase
of the scale in a single transformation. The renormalization group
associates with an initial Boltzmann factor $Z(\phi)$ an orbit
\begin{equation}
Z(\phi,L)=\RG_L Z(\phi),
\label{bolflo}
\end{equation}
parametrized by $L>1$. Due to the semi-group property it
interpolates the sequence
\begin{equation}
\RG_L^n Z(\phi)=\RG_{L^n} Z(\phi),
\end{equation}
obtained by iteration of $\RG_L$. This interpolation is the
motive of the present investigation, in conjunction with the previous
work in \cite{Wi96} on the discrete case. The continuous point of
view has the advantage to allow for infinitesimal renormalization
group transformations, which can be expected close to the identity.

\subsection{Infinitesimal Renormalization Group}

The infinitesimal renormalization group was invented by Wilson
\cite{WK74}. Its importance as {\it exact} dynamical
implementation of scale transformations is receiving increasing
recognition. See also Wegner's review \cite{We76}.

The renormalization group orbit (\ref{bolflo}) satisfies the
remarkable functional differential equation
\begin{equation}
\left\{ L\frac{\partial}{\partial L}-
\left( \DG\phi,\frac{\delta}{\delta \phi}\right)
\right\} Z(\phi,L)=
\frac 12 \left(\frac{\delta}{\delta \phi},\CG
\frac{\delta}{\delta \phi}\right)Z(\phi,L).
\label{rgdgl}
\end{equation}
Here $\DG$ is the generator of dilatations of the field $\phi$.
In real space it is given by
\begin{equation}
\DG\phi(x)=\SG_{L^{-1}}
\left(L\frac{\partial}{\partial L}\SG_L\right)\phi(x)
= \frac{\partial}{\partial L}\SG_L\phi(x)
\biggr\vert_{L=1} =
\left\{ 1-\frac D2 -
x\frac{\partial}{\partial x}\right\}\phi(x).
\label{rgdglz}
\end{equation}
Furthermore, $C$ is a rescaled scale derivative of the fluctuation
covariance. Its explicit expression is
\begin{equation}
C=\SG_{L^{-1}}
\left(L\frac{\partial}{\partial L}\Gamma_L\right)
\SG_{L^{-1}}^T.
\label{contract}
\end{equation}
For a general cutoff function of the form
$\wt{\chi}(p)=F(p^2)$, the covariance (\ref{contract}) is
diagonal in momentum space with eigenvalues
$\wt{C}(p)=-2F^\prime(p^2)$. In the case of the
exponential cutoff (\ref{contract}), it is $C=2\chi$.
In particular, it is independent of $L$ and positive.
The proof of (\ref{rgdglz}) is
\begin{eqnarray}
&& L\frac{\partial}{\partial L}Z(\psi,L) =
L\frac{\partial}{\partial L}\int\dm {\zeta}\; Z(\psi_L+\zeta)
\nonumber \\ && \quad =
\left\{\left(\pl \psi_L,\ppl\right)+
\frac12\left(\ppl,\left(\pl\Gamma_L\right)\ppl\right)\right\}
\int\dm {\zeta} \;Z(\psi_L+\zeta)
\nonumber \\ && \quad =
\left\{\left(\SG_{L^{-1}}\left(\pl \SG_L\right)\psi,\pp\right)+
\frac12\left(\pp,\SG_{L^{-1}}\left(\pl\Gamma_L\right)
\SG_{L^{-1}}^T\pp\right)\right\}
\nonumber \\ && \quad\quad
Z(\psi,L).
\label{rgproof}
\end{eqnarray}
It uses the change of covariance formula for Gaussian measures
and a dilatation identity for functional derivatives. Here
$\psi_L$ stands for $\SG_L\psi$.
The first operator on the right hand side of (\ref{rgproof})
performs an infinitesimal dilatation of the field $\psi$.
Without this dilatation term we would have a functional heat
equation. For the following analysis it is important to rescale
the field and thus carry along the dilatation term.  We should
however concede that the question of ultraviolet and infrared
limit can be approached in a non-rescaled formalism as well.
There the concept of a renormalized trajectory is hidden in
scaling properties of a limit theory.

It is then straight forward to derive a functional differential
equation for the potential $V(\phi,L)=-\log (Z(\phi,L))$. It is
given by
\begin{eqnarray}
&&\left\{
\pl-\left(\DG\phi,\pd\right)
-\frac12\left(\pd,C\pd\right)
\right\}
V(\phi,L)=
\nonumber\\&&\quad
\frac{-1}2\left(\pd V(\phi,L),C\pd V(\phi,L)\right).
\label{rgdglv}
\end{eqnarray}
Here we have collected the linear terms on the left hand side.
Notice that the linearization of (\ref{rgdglv}) at zero
potential coincides with (\ref{rgdglz}). Variations of this
flow equation have proved to be useful both in the study of
perturbation theory and in numerical studies of renormalization
group flows. Polchinski \cite{P84} for instance uses a
flow equation without dilatation term in his beautiful proof of
perturbative renormalizability.

\subsection{Ultraviolet Limit}

Eq. (\ref{flucov}) suggests that we are always performing an
infrared limit upon iteration of (\ref{rgv}). In the rescaled
formalism ultraviolet and infrared limit are closely related.
We include this comment on rescaling to prevent confusion at
this point. Consider a general massless covariance
\begin{equation}
\wt{v}_{\LIR,\LUV}(p)=
\frac 1{p^2}\left\{\wt{\chi}\left(\frac p{\LUV}\right)-
\wt{\chi}\left(\frac p{\LIR}\right)\right\}.
\end{equation}
with two sided cutoffs. The ultraviolet limit refers to sending
the upper cutoff $\LUV$ to $\infty$ in
\begin{equation}
\bar{Z}^{eff}_{\LIR} (\bar{\psi})=
{\cal R}_{\LIR,\LUV} \bar{Z}^{bare}_\LUV (\bar{\psi})=
\int{\rm d}\mu_{v_{\LIR,\LUV}}(\bar{\zeta})
\;\bar{Z}^{bare}_\LUV(\bar{\psi}+\bar{\zeta}),
\end{equation}
keeping the lower cutoff $\LIR$ fixed at a renormalization
scale. Equivalently, the ratio $L=\LUV /\LIR$ is sent to
infinity at fixed lower cutoff $\LIR$. Define a rescaled
bare Boltzmann factor by
\begin{equation}
\bar{Z}^{bare}_\LUV (\bar{\phi})=
Z^{bare}(\SG_\LUV\bar{\phi})=
Z^{bare}(\phi).
\end{equation}
The bare Boltzmann factor is here written in {\it units} of
the ultraviolet scale $\LUV$; to be precise in terms of a
rescaled field (and rescaled couplings). Analogously write the
effective Boltzmann factor in units of the infrared scale $\LIR$,
\begin{equation}
\bar{Z}^{eff}_\LIR (\bar{\psi})=
Z^{eff}(\SG_\LIR\bar{\psi})=
Z^{eff}(\psi).
\end{equation}
Then the renormalization group transformation for the rescaled
quantities is precisely of the form (\ref{rgb}),
\begin{equation}
Z^{eff}(\phi)=\RG_L Z^{bare}(\psi).
\label{dimless}
\end{equation}
Thus keeping the infrared cutoff fixed, the ultraviolet limit is
indeed equivalent with an infrared limit for the rescaled system.
The proof of (\ref{dimless}) is
\begin{eqnarray}
Z^{eff}(\psi)&=& \bar{Z}^{eff}(\SG_{\LIR^{-1}}\psi)
\nonumber \\ &=&
\int {\rm d}\mu_{v_{\LIR,\LUV}}(\bar{\zeta})
\;\bar{Z}^{bare}_\LUV (\SG_{\LIR^{-1}}\psi+\bar{\zeta})
\nonumber \\ &=&
\int {\rm d}\mu_{\SG_{\LUV^{-1}}\Gamma_L\SG_{\LUV^{-1}}^T}
(\bar{\zeta})\; Z^{bare}\left(\SG_{\LUV}( \SG_{\LIR^{-1}}\psi+
\bar{\zeta})\right)
\nonumber \\ &=&
\int\dm{\zeta}\; Z^{bare}(\SG_L\psi+\zeta).
\end{eqnarray}
In practice it is convenient to put $\LIR =1$ in physical
units. Then the outcome of the rescaled renormalization group
is already the desired effective potential. The argument can be
summarized in the following diagram:
\begin{equation*}
\begin{array}{c}
\SG_\LUV\bar{\phi}=\phi \\
\begin{CD}
\bar{Z}^{bare}_\LUV (\bar{\phi})
 @= Z^{bare}(\phi) \\
@V{\RG_{\LIR,\LUV}}VV
 @VV{\RG_L}V \\
\bar{Z}^{eff}_\LIR (\bar{\psi})
 @= Z^{eff}(\psi)
\end{CD} \\
\SG_\LIR\bar{\psi}=\psi
\end{array}
\end{equation*}
In the discrete renormalization group built upon iteration of
(\ref{rgb}), rescaling can be viewed as a stack of these
diagrams on top of each other.

\subsection{Green's Functions}

The effective potential considered here is the generating
function of free propagator amputated connected Euclidean
Green's functions. Let us also include a brief mention of this
property for the sake of completeness. Let
$\bar{W}^{eff}_{\LIR}(\bar{J})$ be the generating
function of the connected Green's functions with free
propagator $v_{\LIR,\LUV}$ and vertices
$\bar{V}^{bare}_{\LUV}(\bar{\phi})$. Then
\begin{equation}
\bar{V}^{eff}_{\LIR}(\bar{\psi})=
\bar{W}^{eff}_{\LIR}(\bar{J})+
\frac 12 \left(\bar{J},v_{\LIR,\LUV}\bar{J}\right),\quad
\bar{\psi}=v_{\LIR,\LUV}\bar{J}.
\label{genfun}
\end{equation}
Eq. (\ref{genfun}) follows from
\begin{gather}
\exp \left\{ -\bar{V}^{eff}_{\LIR} (\bar{\psi}) \right\}=
\int {\rm d} \mu_{v_{\LIR,\LUV}} (\bar{\zeta})
\;\exp \left\{-\bar{V}^{bare}_{\LUV}(\bar{\psi}+\bar{\zeta})\right\}
\nonumber \\
=\exp \left\{ \frac {-1}2
\left( \bar{\psi},v_{\LIR,\LUV}^{-1} \bar{\psi} \right) \right\}
\int {\rm d} \mu_{v_{\LIR,\LUV}} (\bar{\zeta})
\:\exp \left\{ -\bar{V}^{bare}_{\LUV} (\bar{\zeta})
+\left( \bar{\zeta},v_{\LIR,\LUV}^{-1} \bar{\psi} \right)
\right\},
\end{gather}
which uses a shift identity for Gaussian measures. The free
propagator amputated connected Green's functions are given
by the kernels of the development
\begin{equation}
\bar{V}^{eff}_{\LIR}(\bar{\psi})=
\sum_{n=1}^\infty\frac 1{(2n)!}
\int{\rm d}^D\bar{x}_1\ldots{\rm d}^D\bar{x}_{2n}
\;\bar{\psi}(\bar{x})\ldots\bar{\psi}(\bar{x}_{2n})
\;\bar{V}^{eff}_{\LIR,2n}(\bar{x}_1,\ldots,\bar{x}_{2n}),
\label{grepot}
\end{equation}
and the non-amputated connected ones by
\begin{equation}
\bar{W}^{eff}_{\LIR}(\bar{J})=
\sum_{n=1}^\infty\frac 1{(2n)!}
\int{\rm d}^D\bar{x}_1\ldots{\rm d}^D\bar{x}_{2n}
\;\bar{J}(\bar{x})\ldots\bar{J}(\bar{x}_{2n})
\;\bar{W}^{eff}_{\LIR,2n}(\bar{x}_1,\ldots,\bar{x}_{2n}).
\label{gresou}
\end{equation}
The non-amputated connected Green's functions are reconstructed
from the amputated ones through
\begin{gather}
\bar{W}^{eff}_{\LIR,2}(\bar{x}_1,\bar{x}_2)=
-v_{\LIR,\LUV}(\bar{x}_1-\bar{x}_2)+
\nonumber \\
\int {\rm d}^D\bar{y}_1\: {\rm d}^D\bar{y}_2\;
v_{\LIR,\LUV}(\bar{x}_1-\bar{y}_1)
\;v_{\LIR,\LUV}(\bar{x}_2-\bar{y}_2)\;
\bar{V}^{eff}_{\LIR,2}(\bar{y}_1,\bar{y}_2),
\label{aampu}
\end{gather}
and
\begin{gather}
\bar{W}^{eff}_{\LIR,2n}(\bar{x}_1,\ldots,\bar{x}_{2n})=
\nonumber \\
\int {\rm d}^D\bar{y}_1 \ldots {\rm d}^D\bar{y}_{2n}\;
v_{\LIR,\LUV}(\bar{x}_1-\bar{y}_1)\dots
v_{\LIR,\LUV}(\bar{x}_{2n}-\bar{y}_{2n})\;
\bar{V}^{eff}_{\LIR,2}(\bar{y}_1,\ldots,\bar{y}_{2n}).
\label{bampu}
\end{gather}
The connection between the rescaled amputated Green's
functions and the non-rescaled amputated ones on the other
hand is
\begin{equation}
\bar{V}^{ren}_{\LIR}(\bar{\psi})=V^{ren}(\psi),\quad
\SG_{\LIR}\bar{\psi}=\psi.
\end{equation}
The non-rescaled amputated Green's functions are therefore
explicitely given by
\begin{equation}
\bar{V}^{ren}_{\LIR,2n}(\bar{x}_1,\ldots,\bar{x}_{2n})=
\LIR^{n(1+D/2)} V_{2n}(\LIR\bar{x}_1,\ldots,\LIR\bar{x}_{2n})
\label{scalir}
\end{equation}
in terms of the rescaled ones. In the ultraviolet limit
the infrared scale $\LIR$ is kept fixed and (\ref{scalir})
amounts to a finite rescaling. But then the dictionary is
complete. To obtain non-rescaled non-amputated connected
Green's functions from the effective potential one has to
undo rescaling using (\ref{scalir}) and thereafter undo
amputation using (\ref{aampu}) and (\ref{bampu}). It is
clear that the infrared scale $\LIR$ is here an extra piece
of information which has to be supplied from the outside.

\section{Normal Ordering}

We choose to represent the potential in terms of normal
ordered products. The payoff of normal ordering is that
the linear part of the flow equations simplifies to a
scale derivative and a dilatation term. The price to pay
is a more involved bilinear term.

\subsection{Normal Ordering Operator}

We introduce another covariance $v$, which will serve
as normal ordering covariance. Let $v$ be given by
\begin{equation}
\wt{v}(p)=\frac 1{p^2}\wt{\chi}(p),
\label{normal}
\end{equation}
a massless covariance with unit ultraviolet cutoff but
without infrared cutoff. It satisfies
\begin{equation}
v=\SG_{L^{-1}}(v-\Gamma_L)\SG_{L^{-1}}^T.
\end{equation}
As is shown for instance in \cite{Wi96}, this is the
fixed point condition for a flow of normal ordering.
In dimensions $D>2$ the infrared singularity of (\ref{normal})
is integrable. This integrability is sufficient for the below
purposes, in particular for an estimate on the bilinear
renormalization group bracket in eq. (\ref{flopot}). Associated
with (\ref{normal}) is a normal ordering operator acting as
\begin{equation}
:\ZZ (\phi ):_v =
\exp \left \{ \frac{-1}2 \left( \pd,v\pd\right) \right\}
\ZZ (\phi)
\label{order}
\end{equation}
on polynomials, more generally power series, of the field.
The commutator of the normal ordering operator (\ref{order})
with the dilatation operator in the functional differential
equation is computed to
\begin{eqnarray}
&&\exp\left\{\frac{-1}2\left(\pd,v\pd\right)\right\}
\left(\DG\phi,\pd\right)
\exp\left\{\frac12\left(\pd,v\pd\right)\right\}
=\nonumber\\&&\quad
\left(\DG\phi,\pd\right)+
\frac 12\left(\pd,C\pd\right).
\label{commut}
\end{eqnarray}
The intention with the normal ordering covariance (\ref{normal})
was to obtain this identity. Differentiation of (\ref{normal})
with respect to the scale parameter $L$ supplies us with
\begin{equation}
\DG v+v \DG^T=-\SG_{L^{-1}}\left(\pl\Gamma_L\right)\SG_{L^{-1}}^T
=-C.
\end{equation}
From it we conclude that the commutator of the functional
Laplacian, built from the normal ordering covariance, with
the dilatation operator is given by
\begin{equation}
\left[\left(\pd,v\pd\right),\left(\DG\phi,\pd\right)\right]=
\left(\pd,\left(\DG v+v\DG^T\right)\pd\right)=-\left(\pd,C\pd\right).
\end{equation}
But this is the infinitesimal version of (\ref{commut}).
Eq. (\ref{commut}) then follows by integration.

Since the normal ordering covariance (\ref{normal}) is
independent of $L$, the flow equation (\ref{rgdglz}) is
equivalent to
\begin{equation}
\exp\left\{\frac{-1}2\left(\pd,v\pd\right)\right\}
\left\{\pl-\left(\DG\phi,\pd\right)\right\}
\exp\left\{\frac12\left(\pd,v\pd\right)\right\}
Z(\phi,L)=0.
\end{equation}
This equivalence suggests a normal ordered
representation for the Boltzmann factor. We write it in
the form
\begin{equation}
Z(\phi,L)=:\ZZ (\phi,L):_v=
\exp\left\{\frac{-1}2\left(\pd,v\pd\right)\right\}
\ZZ (\phi,L).
\label{norbol}
\end{equation}
We call $\ZZ (\phi,L)$ normal ordered Boltzmann factor.
Strictly speaking it is the pre-image of a normal ordered
Boltzmann factor by the normal ordering operator (\ref{order}).
Therefore it is not decorated with normal ordering colons.
The normal ordered Boltzmann factor then satisfies the first
order functional differential equation
\begin{equation}
\left\{\pl-\left(\DG\phi,\pd\right)\right\}
\ZZ (\phi,L)=0.
\label{scafie}
\end{equation}
A noticeable feature of the normal ordered representation
is that the exact renormalization group equation (\ref{scafie})
exactly performs an infinitesimal scale transformation. It is
equivalent to
\begin{equation}
L\frac{d}{dL}\ZZ(\phi_L,L)=0,
\label{simplez}
\end{equation}
with rescaled field $\phi_L=\SG_L\phi$. In terms of the potential
the theory is not as simple as this due to the non-linearity
of (\ref{rgdglv}). We remark that in perturbation theory we have
in every order to deal with no more than polynomials in $\phi$.
Normal ordering thus requires no more than a finite number of
extra contractions with the normal ordering covariance. They will
be shown to be finite.

\subsection{Homogeneous Solutions}

The scaling fields of the trivial fixed point are polynomial
solutions to (\ref{scafie}). They are given by homogeneous
kernels. Let us have a brief look at them because they will be
used to parametrize the potential. A detailed discussion of
scaling fields can be found for instance in Wegner's review
\cite{We76}. A polynomial
\begin{equation}
\OBS (\phi ,L)=
\frac 1{n!}\int\ddx{n}\;\px{n}\; \OBS_n (x_1,\dots,x_n,L)
\end{equation}
in the field $\phi$ is a solution to the flow equation (\ref{scafie})
iff the kernel satisfies
\begin{eqnarray}
&&
\left\{\pl -\sum_{m=1}^n (\DG^T)^{(m)}\right\}
\OBS_n (x_1,\dots,x_n,L)=
\nonumber\\&&\quad
\left\{\pl -n\left(1+\frac D2\right)-
\sum_{m=1}^n x_m\frac{\partial}{\partial x_m}\right\}
\OBS_n (x_1,\dots,x_n,L)=0.
\label{kersca}
\end{eqnarray}
Eq. (\ref{kersca}) is a homogeneous scaling equation. It is
apparent that the renormalization group flow is a pure scale
transformation in this formulation. The general solution of
(\ref{kersca}) is
\begin{equation}
\OBS_n (x_1,\dots ,x_n,L)=L^{n(1+D/2)}
\OBS_n (Lx_1,\dots ,Lx_n).
\end{equation}
A homogeneous kernel of degree $\kappa$,
$\OBS_n (Lx_1,\dots ,Lx_n)=L^{\kappa}
\OBS_n (x_1,\dots ,x_n),$
yields an eigenvector of the dilatation generator with
eigenvalue $n(1+D/2)+\kappa$. It is called a scaling field.
The eigenvalue is called its real space scaling dimension.
The kernels with negative degree of homogeneity
have powerlike fall-off at large distances and singularities
at small distances. Kernels with positive degree of homogeneity
are non-local and discarded in field theory. A proper
mathematical setting for field theoretic kernels is that of
symmetric distributions on n copies of real space, given by
Fourier integrals
\begin{gather}
\OBS_n (x_1,\dots ,x_n,L)=\nonumber\\
\int\ddp{n}\exp \left(i\sum_{m=1}^n p_m x_m\right)
(2\pi)^D\delta^{(D)} \left(\sum_{m=1}^n p_m\right)
\wt{\OBS}_n (p_1,\dots ,p_{n-1},L).
\label{fourier}
\end{gather}
We will restrict our attention to translation invariant kernels.
The $\delta$-function due to conservation of total momentum is
then conveniently factorized from the momentum space kernels. In
the sequel it is understood that the $n$th momentum of an
$n$-point kernel is $p_n=-\sum_{m=1}^{n-1}p_m$. Furthermore, it
is understood that the momentum space kernels are symmetric
functions of the $n-1$ momenta. Eq. (\ref{fourier})
is a solution to (\ref{scafie}) iff
\begin{equation}
\left\{\pl -\left(D+n\left(1-\frac D2\right)\right)+
\sum_{m=1}^{n-1} p_m\frac{\partial}{\partial p_m}\right\}
\wt{\OBS}_n (p_1,\dots,p_{n-1},L)=0.
\label{momsca}
\end{equation}
The general solution to (\ref{momsca}) is of course given by
\begin{equation}
\wt{\OBS}_n(p_1,\dots,p_{n-1},L)=
L^{D+n(1-D/2)}\OBS \left(\frac{p_1}L,\dots,\frac{p_{n-1}}L\right).
\end{equation}
The power-counting of a homogeneous momentum space kernel of
degree $\wt{\kappa}$ is therefore $D+n(1-D/2)-\wt{\kappa}$.
A momentum derivative reduces the power-counting of a kernel
by one unit. The relevant parts of kernels in field theory
come as sums of scaling fields with positive momentum space
scaling dimension. They are extracted by Taylor expansion in
momentum space. Here field theoretic momentum space kernels will be
required to be symmetric, Euclidean invariant, and regular at
zero momentum.

\subsection{Normal Ordered Potential}

Normal ordering reduces the flow of a Boltzmann factor to
a pure scale transformation. This is not the case for the
potential. Nevertheless it is useful to write potentials
in normal ordered form at least for the case of weak
coupling, where normal ordered products are eigenvectors
of the linearized renormalization group. Let us define
\begin{equation}
V(\phi,L)=:\VV (\phi,L):_v=
\exp\left\{\frac{-1}2\left(\pd,v\pd\right)\right\}
\VV (\phi,L).
\label{norpot}
\end{equation}
in likeness to (\ref{norbol}). This normal ordered potential
then obeys the following non-linear functional differential
equation
\begin{equation}
\left\{\pl-\left(\DG\phi,\pd\right)\right\}\VV (\phi,L)=
-\left\langle \VV(\phi,L),\VV(\phi,L)\right\rangle.
\label{flopot}
\end{equation}
It will be the main dynamical equation of this investigation.
The non-linearity consists of a bilinear renormalization group
bracket
\begin{eqnarray}
&&\left\langle \VV(\phi,L),\VV(\phi,L)\right\rangle=
\nonumber\\&&\quad
\frac 12\left(\frac{\delta}{\delta\phi^{1}},C
\frac{\delta}{\delta\phi^{2}}\right)
\exp\left\{\left(\frac{\delta}{\delta\phi^{1}},
v\frac{\delta}{\delta\phi^{2}}\right)\right\}
\VV(\phi^{1},L) \VV(\phi^{2},L)
\biggr\vert_{\phi^{1}=\phi^{2}=\phi}.
\label{biliterm}
\end{eqnarray}
Here $\phi^{1}$ and $\phi^{2}$ denote two independent
copies of $\phi$. The bilinear term consists of contractions
between two copies of the potential. Notice that every
contraction contains one {\it hard} line $C$ and any
number of {\it soft} lines $v$. Eq. (\ref{flopot}) has
been used before in a proof of Symanzik's improvement program
\cite{Wi88}.

As a byproduct we obtain a normal ordered functional differential
equation for renormalization group fixed points. It is given by
\begin{equation}
\left(\DG\phi,\pd\right)\VV_{*} (\phi)=
\left\langle \VV_{*}(\phi),\VV_{*}(\phi)\right\rangle.
\label{fixpot}
\end{equation}
Renormalization group fixed points are {\it global} solutions
to (\ref{fixpot}). Non-perturbative tools for their investigation
would be a major break through in this theory. Presently our
toolbox contains only the $\epsilon$-expansion and numerical
recipes for truncated systems. An investigation of (\ref{fixpot})
along these lines will be presented elsewhere.

Eq. (\ref{fixpot}) has
a trivial solution $\VV_{*}(\phi)=0$, the free massless
field. In the following we will restrict our attention
to weak perturbations of this trivial fixed point.
It is then appropriate to use iterative methods to solve
(\ref{flopot}).

The proof of (\ref{flopot}) is
\begin{eqnarray}
&&\left\{\pl-\left(\DG\phi,\pd\right)\right\}\VV (\phi,L)=
\nonumber\\&&\quad
\left\{\pl-\left(\DG\phi,\pd\right)\right\}
\exp\left\{\frac 12\left(\pd ,v\pd\right)\right\}V(\phi,L)=
\nonumber\\&&\quad
-\exp\left\{\frac 12\left(\pd ,v\pd\right)\right\}
\frac 12\left(\pd V(\phi,L),C\pd V(\phi,L)\right)=
\nonumber\\&&\quad
-\exp\left\{\frac 12\left(\pd ,v\pd\right)\right\}
\frac 12\left(\pd
\exp\biggl\{\frac 12\left(\pd ,v\pd\right)\right\}\VV (\phi,L),
\nonumber\\&&\quad\quad
C\pd
\exp\left\{\frac 12\left(\pd ,v\pd\right)\biggr\}\VV (\phi,L)
\right)=
\nonumber\\&&\quad
\frac {-1}2\left(\frac{\delta}{\delta\phi^{1}},C
\frac{\delta}{\delta\phi^{2}}\right)
\exp\left\{\frac 12\left(
\frac{\delta}{\delta\phi^{1}}+\frac{\delta}{\delta\phi^{2}},v
\left(\frac{\delta}{\delta\phi^{1}}+\frac{\delta}{\delta\phi^{2}}
\right)\right)\right\}
\nonumber\\&&\quad\quad
\exp\left\{\frac{-1}2\left(\frac{\delta}{\delta\phi^{1}},v
\frac{\delta}{\delta\phi^{1}}\right)\right\}
\exp\left\{\frac{-1}2\left(\frac{\delta}{\delta\phi^{2}},v
\frac{\delta}{\delta\phi^{2}}\right)\right\}
\nonumber\\&&\quad\quad
\VV (\phi^{1},L) \VV (\phi^{2},L)
\biggr\vert_{\phi^{1}=\phi^{2}=\phi}=
\nonumber\\&&\quad
\frac {-1}2\left(\frac{\delta}{\delta\phi^{1}},C
\frac{\delta}{\delta\phi^{2}}\right)
\exp\left\{\left(\frac{\delta}{\delta\phi^{1}},v
\frac{\delta}{\delta\phi^{2}}\right)\right\}
\VV (\phi^{1},L) \VV (\phi^{2},L)
\biggr\vert_{\phi^{1}=\phi^{2}=\phi}
\end{eqnarray}
Undoing the field rescaling (\ref{flopot}), becomes
\begin{equation}
L\frac{{\rm d}}{{\rm d}L}\VV (\phi_L,L)=
-\left\langle \VV(\phi_L,L),\VV(\phi_L,L)\right\rangle.
\label{simplev}
\end{equation}
Since $C$ is positive definite, the left hand side of
(\ref{simplev}) is always negative. It follows that the
scaled potential decreases by value under a scale transformation.
Notice that the difference between the bilinear term
in the non-normal ordered formulation (\ref{rgdglv}) and
the normal ordered one (\ref{flopot}) is a re-normal
ordering operator. In perturbation theory it can be seen
to create additional loops with normal ordering covariance.
Moreover, normal ordering covariances appear only in
contractions of vertices and never at external legs.

\subsection{Bilinear Renormalization Group Bracket}

In the non-linear formulation the bilinear term
is responsible for inhomogeneous terms in scaling
equations to be considered below. Let us write it
explicitely for even monomials $\OBS_{2n}(\phi)$
in the field $\phi$, given by
\begin{equation}
\OBS_{2n} (\phi)=
\frac 1{(2n)!}\int\ddx{2n}\;\px{2n}\;\OBS_{2n} (x_1,\dots,x_{2n}).
\label{eveobs}
\end{equation}
Recall that the bilinear term does not depend on the
scale $L$. The bilinear operation on two monomials
of the form (\ref{eveobs}) can be decomposed into
\begin{equation}
\left\langle \OBS_{2n}(\phi),\OBS_{2m}(\phi)\right\rangle=
\sum_{l=|n-m|}^{n+m-1}
N_{n,m,l} \left( \OBS_{2n}\star\OBS_{2m}\right)_{2l}(\phi),
\label{product}
\end{equation}
and is itself a sum of monomials
\begin{gather}
\left( \OBS_{2n}\star\OBS_{2m}\right)_{2l}(\phi)=
\nonumber \\
\frac 1{(2l)!}\int\ddx{2l}\;\px{2l}\;
\left(\OBS_{2n}\star\OBS_{2m}\right)_{2l} (x_1,\dots,x_{2l}),
\end{gather}
whose kernels are given by a multiple convolutions
\begin{eqnarray}
&&\left(\OBS_{2n}\star\OBS_{2m}\right)_{2l} (x_1,\dots,x_{2l})=
\nonumber\\&&\quad
\frac 1{2(2l)!}
\int{\rm d}y_1\dots{\rm d}y_{2(n+m-l)}
\;C(y_1-y_{n+m-l+1}) \prod_{k=2}^{n+m-l} v(y_k-y_{n+m-l+k})
\nonumber\\&&\quad
\biggl\{\OBS_{2n}(x_1,\dots,x_{n-m+l},y_1,\dots,y_{n+m-l})
\nonumber\\&&\quad\quad
\OBS_{2m}(x_{n-m+l+1},\dots,x_{2l},y_{n+m-l+1},\dots,
y_{2(n+m-l)})+
\nonumber\\&&\quad\quad
((2l)!-1)\; {\rm permutations}\biggr\}.
\label{bilsum}
\end{eqnarray}
The kernels are understood to be symmetric under
permutations of their entries. The multiple convolution
involves one hard propagator $C$ and $n+m-l-1$ soft
propagators $v$, which is at the same time the number of
loops. Furthermore, (\ref{product}) involves a combinatorial
factor
\begin{equation}
N_{n,m,l}=
\frac{(2l)!}
{(n+m-l-1)!(n-m+l)!(m-n+l)!},
\end{equation}
coming from the number of ways in which the contractions
can be made. Eq. (\ref{bilsum}) can be interpreted as
result of the fusion of two vertices. In the process of
fusion links are created, consisting of propagators.

We present an elementary estimate on this fusion product to
get an idea of what kind of analytical properties can be
expected for the effective potentials. The estimate uses an
$L_{\infty,\epsilon}$-norm in momentum space. The estimate
works at this point for $\epsilon \geq 0$, not too large. Later
it will be used for $\epsilon >0$ only. Notice to begin with that
\begin{equation}
\| \wt{C}\|_{\infty,-2\epsilon}, \;
\| \wt{v}\|_{1,-2\epsilon}<\infty,
\end{equation}
for the propagators with exponential cutoff.\footnote{Here
$\| \wt{C}\|_{\infty,-2\epsilon}={\sup }_{p\in\R^D}
\{|\wt{v}(p)| e^{2\epsilon |p|}\} $ and
$\| \wt{v}\|_{1,-2\epsilon} =(2\pi )^{-D}\int{\rm d}^Dp
|\wt{v}(p) |e^{2\epsilon |p|}$
denote the $L_{\infty,-2\epsilon}$- and
$L_{1,-2\epsilon}$-norms in momentum space.}
At $\epsilon =0$ we have for instance $\|\wt{C}\|_\infty=2$ and
$\|\wt{v}\|_\infty\leq 2\pi^{D/2}/(D-2)$. If the Fourier
transformed kernels now satisfy the bounds
\begin{equation}
\| \wt{\OBS}_{2n}\|_{\infty,\epsilon},\;
\| \wt{\OBS}_{2m}\|_{\infty,\epsilon} < \infty,
\end{equation}
that is, are finite in the
$L_{\infty,\epsilon}$-norm,\footnote{The
$L_{\infty,\epsilon}$-norm for the momentum space kernels is
defined as $\| \wt{\OBS}_{2n}\|_{\infty,\epsilon} =
{\rm sup}_{(p_1,\ldots,p_{2n})
\in {\cal P}_{2n}}\{|\wt{\OBS }(p_1,\ldots,p_{2n})|
e^{-\epsilon (|p_1|+\cdots +|p_{2n}|)}\}$ with
${\cal P}_{2n}=\{ (p_1,\ldots ,p_{2n})\in \R^D \times\cdots
\times\R^D | p_1+\cdots +p_{2n}=0\}$ the hyperplane of
total zero momentum.} then it follows that all the
terms (\ref{bilsum}) in the decomposition of the bilinear
operation have finite $L_{\infty,\epsilon}$-norms in momentum
space. They obey
\begin{equation}
\| (\OBS_{2n}\star\OBS_{2m})^\sim_{2l}\|_{\infty,\epsilon}
\leq \frac 12
\;\| \wt{C}\|_{\infty,-2\epsilon}
\;\| \wt{v}\|_{1,-2\epsilon}^{n+m-l-1}
\;\|\wt{\OBS}_{2n}\|_{\infty,\epsilon}
\;\|\wt{\OBS}_{2m}\|_{\infty,\epsilon}.
\end{equation}
Therefore, the renormalization group flow preserves the
$L_{\infty,\epsilon}$-norm of momentum space kernels for
finite scales. It will be shown that the
$L_{\infty,\epsilon}$-norm is also preserved in the
iterative solution of (\ref{basic}). The estimate
immediately follows from the Fourier transform
\begin{eqnarray}
&&\left(\OBS_{2n}\star\OBS_{2m}\right)^\sim_{2l}
(p_1,\dots,p_{2l-1})=
\nonumber\\&&\quad
\frac 1{2(2l)!}
\int\frac{{\rm d}^Dq_1}{(2\pi)^D}\dots
\frac{{\rm d}^Dq_{n+m-l-1}}{(2\pi)^D}
\;\wt{C}(q_{n+m-l})
\;\prod_{k=1}^{n+m-l-1} \wt{v}(q_i)
\nonumber\\&&\quad
\biggl\{\wt{\OBS}_{2n}(p_1,\dots,p_{n-m+l},q_{1},\dots,q_{n+m-l-1})
\nonumber\\&&\quad\quad
\wt{\OBS}_{2m}(p_{n-m+l+1},\dots,p_{2l},-q_{1},\dots,
-q_{n+m-l-1})+
\nonumber\\&&\quad\quad
((2l)!-1)\; {\rm permutations}\biggr\}.
\label{bilsun}
\end{eqnarray}
of (\ref{bilsum}). The $\delta$-functions from translation
invariance have again been removed. The sums of momenta in
the kernels are zero through
\begin{equation}
p_{2l}=-\sum_{m=1}^{2l-1}p_m,\quad
q_{n+m-l}=-\sum_{k=1}^{n-m+l}p_k-\sum_{k=1}^{n+m-l-1}q_k.
\end{equation}
The idea with the parameter $\epsilon$ is to use part
of the exponential large momentum decay of the fluctuation
and normal ordering propagators to compensate a possible large
momentum growth of the kernels. In the initial value problem
for (\ref{flopot}) with $L_\infty$-bounded inital data
this might seem unnecessary. For instance a pure
$\phi^4$-vertex is constant and thus $L_\infty$-bounded.
The evolution preserves $L_\infty$-boundedness for
all {\it finite} scales $L$. However we cannot expect
the solution to be $L_\infty$-bounded uniformly in $L$.
The limit $L\rightarrow\infty$ requires a separate
treatment of zero momentum derivatives and Taylor
remainders of the non-irrelevant kernels. The price to pay
is a growth in momentum space. In the four dimensional
case it is polynomial in powers and logarithms of
momenta. With an exponential bound we are very far on
the safe side.

\section{Perturbation Theory}

In perturbation theory the effective potential comes in form
of a power series
\begin{equation}
\VV (\phi,L,g)=
\sum_{r=1}^\infty \frac{g^r}{r!}
\;\VV^{(r)} (\phi,L)
\label{pertur}
\end{equation}
in a coupling parameter $g$. The effective potential is here
assumed to be zero to zeroth order. The effective potential is
thus expanded around the trivial fixed point. In the sequel
(\ref{pertur}) will be treated as a formal power series in $g$.
The important question of non-summability of (\ref{pertur}) will
not be addressed.

\subsection{Global Perturbation Theory}

We speak of global perturbation theory when the expansion
parameter is independent of the scale $L$. This expansion
is not appropriate in the limit $L\rightarrow\infty$.
Divergent terms appear and call for renormalization.
We nevertheless develop the global expansion to some detail to
see power counting at work. Inserting (\ref{pertur}) into the
bilinear bracket (\ref{biliterm}) and organizing the result again in
powers of $g$ we obtain
\begin{equation}
\left\langle\VV (\phi,L,g),\VV (\phi,L,g)\right\rangle=
\sum_{r=2}^\infty \frac{g^r}{r!} \sum_{s=1}^{r-1}
\begin{pmatrix} r \\ s \end{pmatrix}
\left\langle
\label{kkernel}\VV^{(s)} (\phi,L),
\VV^{(r-s)} (\phi,L)\right\rangle.
\end{equation}
Let us introduce the abbreviation $\KK^{(r)} (\phi,L)$ for
the sum of brackets to order $r$ on the right hand side of
(\ref{kkernel}). The effective potential (\ref{pertur}) is
a solution to the renormalization group equation
(\ref{flopot}) in the sense of a formal power series in $g$
iff it satisfies
\begin{equation}
\left\{\pl-\left(\DG\phi,\pd\right)\right\}\VV^{(r)} (\phi,L)=
-\KK^{(r)} (\phi,L)
\label{perpot}
\end{equation}
to every order $r\geq 1$. Notice that $\KK^{(r)} (\phi,L)$
depends on $\VV^{(s)} (\phi,L)$ to lower orders
$1\leq s\leq r-1$ only. Supplied with initial data,
(\ref{perpot}) can be integrated and yields a recursion
relation for the coefficients in (\ref{pertur}).

\subsubsection{First Order}

To first order (\ref{perpot}) is a homogeneos scaling equation
\begin{equation}
\left\{\pl-\left(\DG\phi,\pd\right)\right\}\VV^{(1)} (\phi,L)=0.
\label{ordone}
\end{equation}
It tells that the evolution is a pure scale transformation to
first order. Consider the case of $\phi^4$-theory. There the
first order is given by a $\phi^4$-vertex
\begin{equation}
\VV^{(1)}(\phi,L)=
\lambda (L)\frac 1{4!}\int {\rm d}^Dx \;\phi (x)^4.
\label{phifour}
\end{equation}
The reader is invited to add a mass and a wave function term if
he wishes. This $\phi^4$-vertex is a solution to the homogeneous
scaling equation (\ref{ordone}) provided that the coupling
flows according to
\begin{equation}
\left\{\frac{{\rm d}}{{\rm d}L}-(4-D)\right\}\lambda (L)=0.
\end{equation}
The first order evolution is therefore
$\lambda (L)=L^{4-D} \lambda (1)$. The conclusion is of course
that the $\phi^4$-vertex is relevant in dimensions $D<4$,
marginal in $D=4$, and irrelevant in $D>4$. In three dimensions
the coupling flows proportional to the scale $L$. It diverges
as $L\rightarrow\infty$. To first order one already sees that
global perturbation theory is unsuitable to perform this limit.

\subsubsection{Second Order}

The inhomogeneous term due to the renormalization group bracket
of two $\phi^4$-vertices is computed to
\begin{gather}
\KK^{(2)}(\phi,L)=
\lambda(L)^2\int {\rm d}^Dx {\rm d}^Dy \biggl\{
\frac 12 \phi(x)\phi(y)\;C(x-y)\;v(x-y)^2+
\nonumber\\
\frac 1{4!} \phi(x)^2\phi(y)^2\;6 C(x-y)\;v(x-y)+
\frac 1{6!} \phi(x)^3\phi(y)^3\;20 C(x-y)\biggr\}.
\label{kktwo}
\end{gather}
Two rather obvious remarks are in place here. First, the bracket
is polynomial if all lower order vertices are polynomials. Second,
the highest power of fields in the bracket is the sum of powers
of the fused vertices minus two. It is instructive to enter the
second order equation (\ref{perpot}) with an ansatz containing
precisely the interactions present in (\ref{kktwo}),
\begin{gather}
\VV^{(2)}(\phi,L)=
\lambda(L)^2\int{\rm d}^Dx{\rm d}^Dy\biggl\{
\frac 12 \phi(x)\phi(y)\;\VV^{(2)}_2(x-y,L)+
\nonumber \\
\frac 1{4!} \phi(x)^2\phi(y)^2\;\VV^{(2)}_4(x-y,L)+
\frac 1{6!} \phi(x)^3\phi(y)^3\;\VV^{(2)}_6(x-y,L)\biggr\}.
\label{anstwp}
\end{gather}
It yields a solution to (\ref{perpot}) iff the kernels in the
ansatz obey the inhomogeneous differential equations
\begin{equation}
\left\{\pl-\sigma^{(2)}_{2n}-\pdx\right\}
\VV^{(2)}_{2n}(x,L)=-\KK^{(2)}_{2n}(x)
\label{inhtwo}
\end{equation}
with
\begin{equation}
\KK^{(2)}_{2}(x)=C(x)v(x)^2,\quad
\KK^{(2)}_{4}(x)=6C(x)v(x),\quad
\KK^{(2)}_{6}(x)=20C(x)
\end{equation}
and
\begin{equation}
\sigma^{(2)}_{2}=3(D-2),\quad
\sigma^{(2)}_{4}=2(D-2),\quad
\sigma^{(2)}_{6}=D-2.
\end{equation}
Eq. (\ref{inhtwo}) is easily integrated. Let us perform the
integration in momentum space and thereby prepare the ground
for the local perturbation expansion below. Fourier transformation
turns (\ref{inhtwo}) into
\begin{equation}
\left\{\pl-\wt{\sigma}^{(2)}_{2n}+\pdp\right\}
\wt{\VV}^{(2)}_{2n}(p,L)=-\wt{\KK}^{(2)}_{2n}(p)
\label{inhtwq}
\end{equation}
with\footnote{The convolution of two momentum space functions
is here defined as
$\wt{F}\star\wt{G}(p)=(2\pi)^{-D}\int{\rm d}^Dq
\wt{F}(p-q)\;\wt{G}(q)$.}
\begin{equation}
\wt{\KK}^{(2)}_{2}(p)=\wt{C}\star\wt{v}\star\wt{v}(p),\quad
\wt{\KK}^{(2)}_{4}(p)=6\wt{C}\star\wt{v}(p),\quad
\wt{\KK}^{(2)}_{6}(p)=20\wt{C}(p),
\end{equation}
and
\begin{equation}
\wt{\sigma}^{(2)}_{2}=2D-6,\quad
\wt{\sigma}^{(2)}_{4}=D-4,\quad
\wt{\sigma}^{(2)}_{6}=-2.\quad
\end{equation}
Notice that our second order ansatz has a prefactor
$\lambda(L)^2$. Therefore the scaling dimensions here are those
of the corresponding monomials relative to the that of two
$\phi^4$-vertices. Another way to write the momentum space
equation (\ref{inhtwq}) is
\begin{equation}
L^{\wt{\sigma}^{(2)}_{2n}}
L\frac{{\rm d}}{{\rm d}L}
\left\{L^{-\wt{\sigma}^{(2)}_{2n}}
\wt{\VV}^{(2)}_{2n}(Lp,L)\right\}=
-\wt{\KK}^{(2)}_{2n}(Lp).
\label{smart}
\end{equation}
In this form it is immediately integrated to
\begin{equation}
\wt{\VV}^{(2)}_{2n}(p,L)=
L^{\wt{\sigma}^{(2)}_{2n}}
\wt{\VV}^{(2)}_{2n}\left(\frac {p}{L},1\right)-
\int_{L^{-1}}^1 \frac{{\rm d}t}{t}
t^{-\wt{\sigma}^{(2)}_{2n}}
\wt{\KK}^{(2)}_{2n}(tp).
\label{pottwo}
\end{equation}
From this expression we can learn that the evolution tends
to forget the irrelevant part of the initial potential as
$L>>1$, whereas the relevant part of the initial potential
is enhanced. We will not renormalize the second order flow
at this instant. But let us remark that due to the damping
of irrelevant initial data an ultraviolet limit depends only
on non-irrelevant interactions in the bare potential.
Furthermore we see that the the integral in (\ref{pottwo})
contains divergent terms in the non-irrelevant case as
$L\rightarrow\infty$. Both powerlike and logarithmic
singularities appear.

A general solution of the second order equation can be
composed of this particular solution and any solution of the
homogeneous scaling equation
\begin{equation}
\left\{\pl-\wt{\sigma}^{(2)}_{2n}+\pdp\right\}
\wt{\VV}^{(2)}_{2n}(p,L)=0.
\label{inhtwx}
\end{equation}
In the sequel we will adopt the following point of view
regarding this freedom. The first order interaction enforces
recursively through the bilinear bracket a certain set of
vertices at higher orders. The homogeneous equation
(\ref{inhtwx}) allows us to introduce further vertices
by hand into the iteration at higher orders, which
are not present in a minimal scheme. We will restrict our
attention to the solution to the renormalization group equation
with a {\it minimal} set of vertices. This solution is
determined by the first order potential. We can think
of the first order as the germ of the theory. Introduction of
other vertices at higher orders might however offer an
interesting way to mix models.

\subsubsection{Higher Orders}

If the first order is polynomial, so are all higher orders in the
minimal scheme. In the case of $\phi^4$-theory the general
form of higher order vertices is
\begin{equation}
\VV^{(r)}(\phi,L)=
\lambda(L)^r
\sum_{n=1}^{r+1}\frac 1{(2n)!}
\int\ddx{2n}\;\px{2n}
\;\VV^{(r)}_{2n}(x_1,\ldots,x_{2n},L).
\label{genfor}
\end{equation}
This general form iterates to every order of perturbation
theory. Other interactions could however be introduced by
hand. The highest connected vertex built from $r$ first
order $\phi^4$-vertices has $2(r+1)$ fields. The renormalization
group equation for the kernels in (\ref{genfor}) reads
\begin{equation}
\left\{\pl -\sigma^{(r)}_{2n}-\sum_{l=1}^{2n} \pdy{l}\right\}
\VV^{(r)}_{2n}(x_1,\ldots,x_{2n},L)=
-\KK^{(r)}_{2n}(x_1,\ldots,x_{2n})
\label{ordrr}
\end{equation}
with real space scaling dimensions
$\sigma^{(r)}_{2n}=m(2+D)-r(4-D)$.
The equivalent equation in momentum space is
\begin{equation}
\left\{\pl -\wt{\sigma}^{(r)}_{2n}+
\sum_{l=1}^{2n-1} \pdq{l}\right\}
\wt{\VV}^{(r)}_{2n}(p_1,\ldots,p_{2n-1},L)=
-\wt{\KK}^{(r)}_{2n}(p_1,\ldots,p_{2n-1})
\label{qrdrr}
\end{equation}
with momentum space scaling dimension
$\wt{\sigma}^{(r)}_{2n}=D+n(2-D)-r(4-D)$.
A compact way of writing (\ref{qrdrr}) then is
\begin{equation}
L^{\wt{\sigma}^{(r)}_{2n}}
L\frac{{\rm d}}{{\rm d}L}
\left\{ L^{-\wt{\sigma}^{(r)}_{2n}}
\wt{\VV}^{(r)}_{2n}(Lp_1,\ldots,Lp_{2n-1},L)\right\}=
-\wt{\KK}^{(r)}_{2n}(Lp_1,\ldots,Lp_{2n-1}),
\end{equation}
which is then integrated to
\begin{gather}
\wt{\VV}^{(r)}_{2n}(p_1,\ldots,p_{2n-1},L)=
\nonumber \\
L^{\wt{\sigma}^{(r)}_{2n}}
\wt{\VV}^{(r)}_{2n}\left(\frac{p_1}L,\ldots,
\frac{p_{2n-1}}L,1\right)
+\int_{L^{-1}}^1\frac{{\rm d}t}t
t^{-\wt{\sigma}^{(r)}_{2n}}
\wt{\KK}^{(r)}_{2n}(tp_1,\ldots,tp_{2n-1}).
\label{high}
\end{gather}
Thereby we have put perturbation theory for the renormalization
group evolution into the form of a recursion relation. A
recursion step consists of all contractions of previous
vertices with one another to a given total order plus
integration of (\ref{high}).

As it stands this perturbation expansion develops singular
coefficients when the scale $L$ is taken to infinity. These
singularities can be removed by renormalization. An elegant way
to renormalize the series goes as follows. In the recursion
formula (\ref{high}) the integral is performed from the
ultraviolet end to the infrared end of the theory. Polchinski
\cite{P84} integrates the non-irrelevant degrees of freedom
precisely the other way using a mixed boundary value problem.
The relevant data is there prescribed on a lower scale than the
irrelevant data. We mention that this approach can be generalized
to a version of Symanzik's improvement program \cite{Wi88}.
This renormalization technology is well developed by now.

In this paper we choose another route. We will search (and find) a
way to formulate the theory in terms of quantities which are
{\it independent} of the scale $L$. These quantities are
auto-renormalized. They are identical with their scaling limits.
The idea is switch to another form of perturbation theory.

\subsection{Local Perturbation Theory}

We speak of local perturbation theory when the expansion parameter
$g$ is taken to depend on the scale $L$. A particular form of
local perturbation theory is the running coupling expansion for
renormalized trajectories proposed in \cite{Wi96}. There the
ambition is to find renormalization group flows whose scale
dependence comes exclusively in form of a running coupling $g(L)$.
The expansion coefficients themselves are scale independent.
We will develop such an expansion for the $\phi^4$-trajectory
in the infinitesimal renormalization group setup. A forerunner
with discrete renormalization group transformation is found in
\cite{Wi96}. We intend to solve the flow equation (\ref{flopot})
in terms of a power series
\begin{equation}
\VV (\phi, g(L))=
\sum_{r=1}^\infty \frac{g(L)^r}{r!} \VV^{(r)}(\phi).
\label{scaans}
\end{equation}
The interaction $\VV^{(r)}(\phi)$ is assumed to be independent
of $L$. We choose the $\phi^4$-coupling as expansion parameter.
It is defined through the condition
\begin{equation}
\wt{\VV}_{4}(0,0,0,g(L))=g(L)
\label{coupling}
\end{equation}
on the four point kernel at zero momentum. To be precise, we
impose the perturbative constraint that (\ref{coupling}) be
zero for all orders larger than one. The $\phi^4$-coupling is
by no means the only possible choice in this approach. It
is however a natural candidate when dealing with $\phi^4$-theory.
It will be used in the following. In order to solve (\ref{flopot})
in terms of (\ref{scaans}) we also require another power series
\begin{equation}
L\frac{{\rm d}}{{\rm d}L}g(L)=
\beta\left(g(L)\right)=
\sum_{r=1}^\infty \frac{g(L)^r}{r!}\beta^{(r)}.
\label{betafun}
\end{equation}
The outcome of this analysis is a curve in the space of effective
interactions parametrized by $g$. The $\beta$-function
(\ref{betafun}) says how the renormalization group acts on
interactions on this curve. The coefficients $\beta^{(r)}$ form
a second set of unknowns, besides those in $\VV^{(r)}(\phi)$. They
have to come out of the theory. To be more precise, the
$\beta$-function substitutes those degrees of freedom which are
removed by (\ref{coupling}). The intent with (\ref{scaans}) is
a renormalization group orbit which is {\it not} parametrized
by $L$ but rather by $g$. As we will see such a curve is
indeed determined to all orders of perturbation theory once
we supply an appropriate first order interaction. In the
case of $\phi^4$-theory the appropriate first order interaction
is
\begin{equation}
\VV^{(1)}(\phi)=\frac 1{4!}\int{\rm d}^Dx \;\phi(x)^4,
\label{firstord}
\end{equation}
a $\phi^4$-vertex. The corresponding renormalization group
orbit is called the $\phi^4$-trajectory. It is the object of
principle interest in massless $\phi^4$-theory. Potentials
on the $\phi^4$-trajectory are said to scale. The first
order interaction (\ref{firstord}) turns out to require
a slight modification in four dimensions. We need to
add also a wave function term. We will do this below.

\subsubsection{Scaling Equations}

We insert the expansion (\ref{scaans}) into the renormalization
group equation (\ref{flopot}) and deduce a system of equations
for the unknowns therefrom. This system is called the set of
scaling equations. We have
\begin{equation}
\pl \VV(\phi,g(L))=
\beta\left(g(L)\right)\frac{\partial}{\partial g(L)}
\VV(\phi,g(L))=
\sum_{r=1}^\infty\frac{g(L)^r}{r!}
\sum_{s=1}^{r}\begin{pmatrix}r\\s\end{pmatrix}
\beta^{(s)}\VV^{(r-s+1)}(\phi).
\end{equation}
Therefore, our local perturbation expansion is a solution to
(\ref{flopot}) in the sense of a formal power series iff
the unknowns obey
\begin{equation}
\sum_{s=1}^r\begin{pmatrix}r\\s\end{pmatrix}
\beta^{(s)}\VV^{(r-s+1)}(\phi)-
\left(\DG\phi,\pd\right)\VV^{(r)}(\phi)=
-\sum_{s=1}^{r-1}\begin{pmatrix}r\\s\end{pmatrix}
\left\langle V^{(s)}(\phi),\VV^{(r-s)}(\phi)\right\rangle
\label{scaling}
\end{equation}
to all orders $r\geq 1$. The most remarkable property of
(\ref{scaling}) is again independence of $L$.
Another way to think of (\ref{scaling}) is that we are
looking for a fixed point of the transformation composed
of a renormalization group step and a transformation of
the four point coupling. We organize (\ref{scaling})
into a recursion relation.

\subsubsection{First Order}

Scaling requires of $\VV^{(1)}(\phi)$ to be a scaling field of
the trivial fixed point. To first order (\ref{scaling}) reads
\begin{equation}
\left\{\beta^{(1)}-\left(\DG\phi,\pd\right)\right\}
\VV^{(1)}(\phi)=0.
\label{orone}
\end{equation}
It requires of $\VV^{(1)}(\phi)$ to be an eigenvector of the
generator of dilatations, and of $\beta^{(1)}$ to be its eigenvalue.
The $\phi^4$-vertex (\ref{firstord}) is indeed an eigenvector.
Its eigenvalue has the familiar value
\begin{equation}
\beta^{(1)}=4-D.
\label{eigone}
\end{equation}
Here three remarks are in place. First, if we want to add further
terms to the first order (\ref{firstord}) then (\ref{orone})
requires of all other terms to be eigenvectors with this same
eigenvalue (\ref{eigone}). In four dimensions the eigenvalue
(\ref{eigone}) is zero. There one has indeed another marginal
eigenvector, the wave function term. Second, we could have
considered any other scaling field of the trivial fixed point as
our first order starting point, for instance a $\phi^6$-vertex
in three dimensions. All of the scaling fields generate interesting
trajectories, for which this theory equally well applies. Third,
the eigenvalue does not need to be larger or equal to zero. We can
as well start with an irrelevant scaling field. In fact the
$\phi^4$-vertex {\it is} marginally irrelvant in four dimensions.

\subsubsection{Second Order}

The first order equation (\ref{orone}) is special in that it is
homogeneous. All higher order scaling equations are inhomogeneous.
The second order equation is given by
\begin{equation}
\left\{2\beta^{(1)}-\left(\DG\phi,\pd\right)\right\}
\VV^{(2)}(\phi)=
-\beta^{(2)}\VV^{(1)}(\phi)-
2\left\langle\VV^{(1)},\VV^{(1)}\right\rangle.
\label{ortwo}
\end{equation}
Notice that unlike in the global expansion there is as well
an inhomogeneous term due to the flow of the coupling parameter.
Minus the right hand side of (\ref{ortwo}) is explicitely
computed to
\begin{gather}
\KK^{(2)}(\phi)=
\frac {\beta^{(2)}}{4!}\int{\rm d}^Dx\phi(x)^4+
\int{\rm d}^Dx{\rm d}^Dy\biggl\{
\frac 12\phi(x)\phi(y) C(x-y)v(x-y)^2+
\nonumber\\
\frac 1{4!} \phi(x)^2\phi(y)^2 6C(x-y)v(x-y)+
\frac 1{6!} \phi(x)^3\phi(y)^3 20 C(x-y) \biggr\}.
\label{inhomtwo}
\end{gather}
The general solution of (\ref{ortwo}) consists of
special solution plus any solution of the homogeneous
equation. In the global expansion special solutions were
singled out by boundary conditions. Here we will do something
different. We will only admit solutions which are given by
finite kernels in momentum space. Furthermore, we will restrict
our attention to local interactions in the minimal scheme.
It then turns out that there exists only one such solution. This
special solution contains precisely those vertices, which are
enforced by the first order interaction. Therefore, we make the
following ansatz
\begin{gather}
\VV^{(2)}(\phi)=
\sum_{n=1}^{3}\frac 1{(2n)!}\int\ddx{2n}\;\px{2n}
\;\VV^{(2)}_{2n}(x_1,\ldots,x_{2n}).
\label{anstwo}
\end{gather}
From (\ref{inhomtwo}) it is clear that we can only succeed in
terms of distributional kernels. We therefore Fourier transform
(\ref{ortwo}) and thereby obtain
\begin{equation}
\left\{2\beta^{(1)}-(D-n(D-2))+
\sum_{m=1}^{2n-1}\pdq{m}\right\}
\wt{\VV}^{(2)}_{2n}(p_1,\ldots,p_{2n-1})=
-\wt{\KK}^{(2)}_{2n}(p_1,\ldots,p_{2n-1})
\label{pdetwo}
\end{equation}
with inhomogeneous terms given by
\begin{gather}
\wt{\KK}^{(2)}_2(p)=
\wt{C}\star\wt{v}\star\wt{v}(p),
\\
\wt{\KK}^{(2)}_4(p_1,p_2,p_3)=
\beta^{(2)}+6\wt{C}\star\wt{v}(p_1+p_2),
\\
\wt{\KK}^{(2)}_6(p_1,\ldots,p_5)=
20\wt{C}(p_1+p_2+p_3).
\label{kscatwo}
\end{gather}
The right hand sides are here understood to be symmetrized in
the momenta. We do not write this symmetrization explicitely
in order to simplify the notation. Here we recognize again the
the second order scaling dimension $\wt{\sigma}^{(2)}_{2n}=
D-n(D-2)-2\beta^{(1)}$, which was introduced above. The
general method to solve the differential equation (\ref{pdetwo})
is explained below. We immediately apply it to this second order
equation. Consider first the six point interaction. Since
$\wt{\sigma}^{(2)}_6=-2$, it is irrelevant (in any dimension),
we can immediately integrate the six point interaction
to\footnote{It is instructive to perform this integral. The result
is $\wt{\VV}^{(2)}_6(p_1,\ldots,p_5)=10(p_1+p_2+p_3)^{-2}
\left\{\exp\left(-(p_1+p_2+p_3)^2\right)-1\right\}$. This
expression is regular at zero momentum and a bounded
function of the momenta. It is of the form of a cutoff
propagator.}
\begin{equation}
\wt{\VV}^{(2)}_6(p_1,\ldots,p_5)=
-\int_{0}^{1}\frac{{\rm d}t}{t} t^{2}
\wt{\KK}^{(2)}_6(tp_1,\ldots,tp_5).
\label{sixtwo}
\end{equation}
Notice the similarity with the analogous expression in the
global expansion. Formally it can be obtained by taking $L$ to
infinity in (\ref{high}). We then treat the four point
interaction. It is computed in two parts. The first part concerns
its value at zero momentum. By definition of our expansion
parameter we have to satisfy
\begin{equation}
0=\beta^{(2)}+6\wt{C}\star\wt{v}(0).
\label{betatwo}
\end{equation}
Using the $\phi^4$-coupling to organize the expansion, we
imposed that $\wt{\VV}^{(2)}_4(0,0,0)$, and all
higher orders of the four point kernel at zero momentum, be
zero. Eq. (\ref{betatwo}) determines the value of $\beta^{(2)}$.
In four dimensions its value is computed to
\begin{equation}
\beta^{(2)}=\frac{-6}{(4\pi)^2}.
\end{equation}
As a consequence the $\phi^4$-vertex is indeed marginally
irrelevant in four dimensions. Thereby the renormalization
group flow on the $\phi^4$-trajectory is not asymptotically
free in the ultraviolet. Nevertheless the $\phi^4$-trajectory
is a well defined object at weak coupling. The irrelevant
remainder of the four point vertex is then integrated to
\begin{equation}
\wt{\VV}^{(2)}_4(p_1,p_2,p_3)=
-6\int_0^1\frac{{\rm dt}}{t}t^{4-D}
\left\{\wt{C}\star\wt{v}(tp_1+tp_2)-
\wt{C}\star\wt{v}(0)\right\}.
\label{fourtwo}
\end{equation}
Here it becomes transparent how the flow of the coupling
parameter saves us from a logarithmic singularity in four
dimensions. Notice also that the subtraction of the
zero momentum piece is unnecessary in three dimensions.
There this vertex is already irrelevant. The two point
kernel finally requires most attention and poses even an
obstacle at this stage. Due to Euclidean invariance we
have
\begin{equation}
\wt{\VV}^{(2)}_2(p)=A(p^2),\quad
\wt{\KK}^{(2)}_2(p)=2B(p^2).
\label{invtwo}
\end{equation}
It is convenient to trade $p^2$ for a new variable $u$. The
scaling dimension of the two point kernel is
$\wt{\sigma}^{(2)}_2=2D-6$, and is two in four dimensions.
The scaling equation for the two point kernel then becomes
\begin{equation}
\left\{u\frac{{\rm d}}{{\rm d}u}-1\right\} A(u)= -B(u).
\label{scamas}
\end{equation}
Its solution requires a second order Taylor expansion with
remainder term for the function
\begin{equation}
A(u)=
A(0)+u\;A^{\prime}(0)+
\frac{u^2}2 \int_0^1{\rm d}t (1-t) A^{\prime\prime}(tu).
\label{taylortwo}
\end{equation}
The zero momentum value is directly determined. Its value is
\begin{equation}
A(0)=B(0)=\frac 12\wt{C}\star\wt{v}\star\wt{v}(0)=
\frac 1{(4\pi)^4}\left(2\log (2)-\log (3)\right),\quad D=4.
\label{masstwo}
\end{equation}
It can be interpreted as a second order mass parameter. The
Taylor remainder in (\ref{taylortwo}) is computed from the
second derivative of (\ref{scamas}). We have
\begin{equation}
\left\{u\frac{{\rm d}}{{\rm d}u}+1\right\}
A^{\prime\prime}(u)= -B^{\prime\prime}(u).
\label{scaren}
\end{equation}
Two $u$-derivatives have changed the powercounting by four units.
As a consequence the second $u$-derivative is irrelevant and is
therefore integrated to
\begin{equation}
A^{\prime\prime}(u)=
-\int_0^1{\rm d}t\;B^{\prime\prime}(tu)
\end{equation}
in complete analogy to the case of the six point
kernel.\footnote{
A parameter representation for the inhomogeneous mass term is
$\wt{\KK}^{(2)}_2(p)=\wt{C}\star\wt{v}\star\wt{v}(p)=
2(4\pi)^{-D}\int_1^\infty{\rm d}\alpha_1\int_1^\infty{\rm d}\alpha_2
(\alpha_1\alpha_2+\alpha_1+\alpha_2)^{-D/2}
\exp\left(-\alpha_1\alpha_2 p^2/
(\alpha_1\alpha_2+\alpha_1+\alpha_2)\right)$.
Notice that it is regular function of the variable $p^2$ in
four dimensions.}
The grain of salt is the first $u$-derivative,
a wave function term. Its scaling equation is
\begin{equation}
u\frac{{\rm d}}{{\rm d}u}A^{\prime}(u)= -B^{\prime}(u).
\label{scarem}
\end{equation}
Since it is marginal, it does not determine the first $u$-derivative
at zero momentum. Furthermore, it requires
\begin{equation}
B^{\prime}(0)=0.
\label{nowave}
\end{equation}
Eq. (\ref{nowave}) can be checked to be false in four dimensions.
The problem is a wave function term which is generated dynamically
to second order. As it stands the theory is inconsistent. This
problem is solved by introduction of a wave function term to first
order. Thus we extend (\ref{firstord}) to
\begin{equation}
\VV^{(1)}(\phi)=
\frac 1{4!}\int{\rm d}^Dx \phi(x)^4+
\frac {\zeta^{(1)}}2 \int{\rm d}^Dx
\partial_\mu\phi(x)  \partial_\mu\phi(x).
\label{firsttwo}
\end{equation}
A first order wave function term is consistent with
(\ref{orone}) in four dimensions. Both terms in (\ref{firsttwo})
are marginal. The first order scaling equation leaves the new
parameter $\zeta^{(1)}$ undetermined. We use this freedom to
satisfy the second order constraint (\ref{nowave}). It is a
general feature of this approach that marginal parameters
are determined one order later than the other ones to a given
order. Nevertheless we have a recursive perturbation theory.
It is now consistent to all orders. Having accompanied the
$\phi^4$-vertex with a first order wave function term, we find
a few more effective interactions in the analogue of
(\ref{inhomtwo}). With (\ref{firstord}) replaced by
(\ref{firsttwo}), it becomes
\begin{gather}
K^{(2)}(\phi)=
\frac {\beta^{(2)}}{4!}\int{\rm d}^Dx\phi(x)^4+
\int{\rm d}^Dx{\rm d}^Dy\biggl\{
\frac 12\phi(x)\phi(y) C(x-y)v(x-y)^2+
\nonumber\\
\frac 1{4!} \phi(x)^2\phi(y)^2 6C(x-y)v(x-y)+
\frac 1{6!} \phi(x)^3\phi(y)^3 20 C(x-y) \biggr\}+
\nonumber\\
\frac {\beta^{(2)}\zeta^{(1)}}2\int{\rm d}^Dx
\partial_\mu\phi(x)\partial_\mu\phi(x)+
\int{\rm d}^Dx{\rm d}^Dy\biggl\{
\frac {(\zeta^{(1)})^2}2\phi(x)\phi(y)
2(-\bigtriangleup_x)(-\bigtriangleup_y)C(x-y)+
\nonumber\\
\frac{\zeta^{(1)}}{2}\phi(x)^2
2(-\bigtriangleup_y)C(y) v(y)+
\frac {\zeta^{(1)}}{4!} \phi(x)\phi(y)^3
8(-\bigtriangleup_y)C(y).
\label{inhomnew}
\end{gather}
It follows that we have to replace eq. (\ref{kscatwo}) by
\begin{gather}
\wt{\KK}^{(2)}_2(p)=
\wt{C}\star\wt{v}\star\wt{v}(p)+
\zeta^{(1)} a+\beta^{(2)}\zeta^{(1)}p^2+
2(\zeta^{(1)})^2 (p^2)^2 \wt{C}(p),
\\
\wt{\KK}^{(2)}_4(p_1,p_2,p_3)=
\beta^{(2)}+6\wt{C}\star\wt{v}(p_1+p_2)+
8 \zeta^{(1)}\;p_1^2 \wt{C}(p_1),
\\
\wt{\KK}^{(2)}_6(p_1,\ldots,p_5)=
20\wt{C}(p_1+p_2+p_3).
\label{kscanew}
\end{gather}
The constant $a$ is here given by a the convergent one loop
integral
\begin{equation}
a=2\int{\rm d}^Dy (-\bigtriangleup_y)C(y)v(y)
=4\int\frac{{\rm d}^Dp}{(2\pi)^D}\exp (-2p^2)
=(4\pi)^{-2},\quad D=4.
\label{constant}
\end{equation}
We then proceed exactly as above. The new terms in (\ref{kscanew})
do not disturb the recursion. The six point kernel is not
affected at all. Neither is the four point kernel at zero
momentum. Therefore the coefficent $\beta^{(2)}$ is independent
of the new constant. The quartic remainder gets an extra contribution
and requires first knowledge of $\zeta^{(1)}$. Spelled out explicitely,
the eq. (\ref{nowave}) becomes
\begin{equation}
0=\beta^{(2)}\zeta^{(1)}+
\frac{\partial}{\partial(p^2)}
\wt{C}\star\wt{v}\star\wt{v}(p)\biggr\vert_{p=0}.
\label{waveone}
\end{equation}
It determines the first order wave function parameter. The second
order work is therefore best organized as follows. We first compute
$\beta^{(2)}$ from (\ref{betatwo}), and second $\zeta^{(1)}$ from
(\ref{waveone}). Knowing of the marginal data we then compute
all kernels in terms of their Taylor expansions and remainders.
In the normal ordered formulation the order in which the kernels
are computed is of no importance. The value of the first order
wave function paramter comes out as
\begin{equation}
\zeta^{(1)}=\frac{-1}{18(4\pi)^2}.
\end{equation}
The second order effective mass parameter is then changed from
(\ref{masstwo}) to
\begin{equation}
A(0)=
\frac 1{(4\pi)^4}\left(2\log (2)-\log (3)-\frac 1{36}\right).
\label{massnew}
\end{equation}
The changed remainder terms of the quadratic and the quartic
kernel follow immediately from (\ref{kscanew}). This completes
the calculation of the second order. In summary we have
extracted the non-irrelevant part by Taylor expansion. The
Taylor coefficients have been determined directly by evaluation
at zero momentum. The irrelevant part, including the Taylor
remainders, have been obtained by integration of the corresponding
scaling equations.

\subsubsection{Higher Orders}

The second order scheme generalizes to third and higher orders.
Assume that we have computed $\VV^{(s)}(\phi)$ and $\beta^{(s)}$
to all orders $1\leq s\leq r-1$ except for $\zeta^{(r-1)}$.
Then we first compute $\beta^{(r)}$, thereafter $\zeta^{(r-1)}$,
and then the $\VV^{(r)}(\phi)$ except for $\zeta^{(r)}$. The
non-irrelevant degrees of freedom are again separated out by
Taylor expansion in momentum space. A convenient notation for
this Taylor expansion goes as follows. Introduce scaling fields
\begin{gather}
\OBS_{2,0}(\phi)=\frac 12\int{\rm d}^Dx\;\phi(x)^2, \quad
\OBS_{2,2}(\phi)=\frac 12\int{\rm d}^Dx\;
\partial_\mu\phi(x) \partial_\mu\phi(x), \\
\OBS_{4,0}(\phi)=\frac 1{4!}\int{\rm d}^Dx\;\phi(x)^4,
\end{gather}
with scaling dimensions
\begin{gather}
\left(\DG\phi,\pd\right)\OBS_{2,0}(\phi)=
2\OBS_{2,0}(\phi),\quad
\left(\DG\phi,\pd\right)\OBS_{2,2}(\phi)=0, \\
\left(\DG\phi,\pd\right)\OBS_{4,0}(\phi)=
(4-D)\OBS_{2,0}(\phi).
\label{scalingdims}
\end{gather}
In four dimensions, the non-irrelevant part of the effective
potential can be written as a sum
\begin{equation}
\VV^{(r)}_{rel}(\phi)=
\mu^{(r)}\OBS_{2,0}(\phi)+
\zeta^{(r)}\OBS_{2,2}(\phi)+
\lambda^{(r)}\OBS_{4,0}(\phi).
\end{equation}
The general form of the effective potential to order $r$ in
the minimal scheme is
\begin{equation}
\VV^{(r)}(\phi)=
\sum_{n=1}^{r+1}\frac{1}{(2n)!}
\int\ddx{2n}\;\px{2n}\;\VV^{(r)}_{2n}(x_1,\ldots,x_{2n}).
\label{orderrsca}
\end{equation}
The non-irrelevant coupling constants are, in terms of these
kernels, given by
\begin{gather}
\mu^{(r)}=
\int{\rm d}^Dx_2 \;\VV^{(r)}_2(x_1,x_2)=
\wt{\VV}^{(r)}_2(0), \\
\zeta^{(r)}=
\frac{-1}{2D}\int{\rm d}^Dx_2\; (x_1-x_2)^2\;\VV^{(r)}(x_1,x_2)=
\frac{\partial}{\partial (p^2)} \wt{\VV}^{(r)}_2(p)
\biggr\vert_{p=0}, \\
\lambda^{(r)}=
\int{\rm d}^Dx_2\;{\rm d}^Dx_3\;{\rm d}^Dx_4\;
\VV^{(r)}_4(x_1,x_2,x_3,x_4)=
\wt{\VV}^{(r)}_4(0,0,0).
\end{gather}
A compact notation for their extraction from the polynomial
(\ref{orderrsca}) is the formal pairing
\begin{equation}
\mu^{(r)}=\left(\OBS_{2,0},\VV^{(r)}\right),\quad
\zeta^{(r)}=\left(\OBS_{2,2},\VV^{(r)}\right),\quad
\lambda^{(r)}=\left(\OBS_{4,0},\VV^{(r)}\right).
\end{equation}
Its meaning is, as was said before, nothing but Taylor expansion
in momentum space. The scaling equation to order $r$ can be put
into the form
\begin{equation}
\left\{r\beta^{(1)}-\left(\DG\phi,\pd\right) \right\}
\VV^{(r)}(\phi)=-\KK^{(r)}(\phi).
\label{scalingr}
\end{equation}
The right hand side of (\ref{scalingr}) still depends on
$\beta^{(r)}$ and $\zeta^{(r-1)}$. It is given by
\begin{equation}
\KK^{(r)}(\phi)=
\sum_{s=2}^{r}\begin{pmatrix}r\\s\end{pmatrix}
\beta^{(s)}\VV^{(r-s+1)}(\phi)+
\sum_{s=1}^{r-1}\begin{pmatrix}r\\s\end{pmatrix}
\left\langle\VV^{(s)},\VV^{(r-s)}\right\rangle.
\label{inhomor}
\end{equation}
Projecting both sides of (\ref{scalingr}) to $\OBS_{4,0}(\phi)$
we find
\begin{equation}
(r-1)\beta^{(1)}\lambda^{(r)}=
-\beta^{(r)}-\sum_{s=2}^{r-1}
\begin{pmatrix}r\\s\end{pmatrix}
\beta^{(s)}\lambda^{(r-s+1)}-
\sum_{s=1}^{r-1}
\begin{pmatrix}r\\s\end{pmatrix}
\left(\OBS_{4,0},
\left\langle\VV^{(s)},\VV^{(r-s)}\right\rangle\right).
\label{detofbeta}
\end{equation}
Since $\lambda^{(r)}=0$ by construction, eq. (\ref{detofbeta})
determines the coefficient $\beta^{(r)}$. The right hand side of
(\ref{detofbeta}) does not depend on $\zeta^{(r-1)}$ since
$\left(\OBS_{4,0},\left\langle \VV^{(1)}(\phi),\OBS_{2,2}
\right\rangle\right)=0$. Thus the order $r-1$ wave function term
does not contribute through the renormalization group bracket
to an effective $\phi^4$-vertex. Thereafter projecting both sides
of (\ref{scalingr}) to $\OBS_{2,2}(\phi)$ we find
\begin{equation}
r\beta^{(1)}\zeta^{(r)}=
-\begin{pmatrix}r\\ 2\end{pmatrix}
\beta^{(2)}\zeta^{(r-1)}-
\sum_{s=3}^{r}\begin{pmatrix}r\\s\end{pmatrix}
\beta^{(s)}\zeta^{(r-s+1)}-
\sum_{s=1}^{r-1}\begin{pmatrix}r\\s\end{pmatrix}
\left(\OBS_{2,2},\left\langle\VV^{(s)},\VV^{(r-s)}
\right\rangle\right).
\label{detofzeta}
\end{equation}
In four dimensions the left hand side is zero since
$\beta^{(1)}=4-D=0$. But then (\ref{detofzeta}) determines the
value of $\zeta^{(r-1)}$. Notice that $\beta^{(2)}$ is not
zero. Notice further that the renormalization group bracket
does not depend on $\zeta^{(r-1)}$ because $\left(\OBS_{2,2},
\left\langle\VV^{(1)},\OBS_{2,2}\right\rangle\right)=0$.
The reason is that the covariance $C$ is regular at
zero momentum. The rest of the work is immediately put to order.
The effective mass parameter to order $r$ follows from
\begin{equation}
\left(r\beta^{(1)}-2\right)\mu^{(r)}=
-\sum_{s=2}^{r}\begin{pmatrix}r\\s\end{pmatrix}
\beta^{(s)}\mu^{(r-s+1)}-
\sum_{s=1}^{r-1}\begin{pmatrix}r\\s\end{pmatrix}
\left(\OBS_{2,0},\left\langle\VV^{(s)},
\VV^{(r-s)}\right\rangle\right).
\end{equation}
Its computation requires the extraction of the effective mass
term in the renormalization group bracket. The computation of
$\VV^{(r)}_{rel}(\phi)$ except for $\zeta^{(r)}$ is complete.
The irrelevant part is directly integrated following exactly the
scheme of the second order calculation. Eq. (\ref{inhomor}) can
be expanded into a polynomial
\begin{equation}
\KK^{(r)}(\phi)=
\sum_{n=1}^{r+1}\frac{1}{(2n)!}
\int\ddx{2n}\;\px{2n}\;\KK^{(2r)}(x_1,\ldots,x_{2n}).
\end{equation}
The scaling equations for the Fourier transformed kernels are
given by
\begin{equation}
\left\{-\wt{\sigma}^{(r)}_{2n}+
\sum_{m=1}^{2n-1}p_m\frac{\partial}{\partial p_m}\right\}
\wt{\VV}^{(r)}_{2n}(p_1,\ldots,p_{2n-1})=
-\wt{\KK}^{(r)}_{2n}(p_1,\ldots,p_{2n-1}).
\end{equation}
For $n\geq 3$, the scaling dimension is
$\wt{\sigma}^{(r)}_{2n}=D+n(2-D)-r(4-D)=4-2n <0$, $D=4$.
Those kernels are therefore all irrelevant. The scaling
equation is in this case integrated to
\begin{equation}
\wt{\VV}^{(r)}_{2n}(p_1,\ldots,p_{2n-1})=
-\int_0^1\frac{{\rm d}t}{t}t^{-\wt{\sigma}^{(r)}_{2n}}
\;\wt{\KK}^{(r)}_{2n}(tp_1,\ldots,tp_{2n-1}).
\end{equation}
The integral converges because of the negative power counting.
The zero momentum part of the four point kernel has already
successfully been transfered to the $\beta$-function. Its
remainder is irrelevant and integrated to
\begin{equation}
\wt{\VV}^{(r)}_{4}(p_1,p_2,p_3)=
-\int_0^1\frac{{\rm d}t}{t}
\left\{\wt{\KK}^{(r)}_4(tp_1,tp_2,tp_3)-
\wt{\KK}^{(r)}_4(0,0,0)\right\}.
\end{equation}
The integral converges due to the subtraction at zero momentum.
Finally, the two point kernel is reconstructed with the help of
\begin{equation}
\wt{\VV}^{(r)}_2(p)=A(p^2),\quad
\wt{\KK}^{(r)}_2(p)=2B(p^2),
\end{equation}
and
\begin{equation}
A(u)=\mu^{(r)}+\zeta^{(r)}u+
\frac{u^2}{2}\int_0^1{\rm d}s(1-s)A^{\prime\prime}(su)
\end{equation}
through
\begin{equation}
A^{\prime\prime}(u)=
-\int_0^1 {\rm d}t B^{\prime\prime}(tu).
\end{equation}
The scheme is now complete. We have a manifestly finite recursive
local perturbation theory.

\section{Regularity}

The integration of renormalization group differential
equations generally requires initial data. In the local
expansion, where we no more have evolution equations, we
substitute the initial data by a requirement that the
solutions be finite and regular.

\subsection{Renormalization Group PDEs}

In the perturbation expansion the recursion to each higher
order consists of solving a set of renormalization group
PDEs of the general form
\begin{equation}
\left\{p\frac{\partial}{\partial p}-\sigma\right\}
F(p)=G(p).
\label{pde}
\end{equation}
Here $\sigma\in\Z$, and $G(p)$ is a given function of
$p\in\R^N$. In this section we will have a look at the
{\it regular} solutions of (\ref{pde}).

\subsubsection{Irrelevant Case}

The irrelevant case is defined by $\sigma <0$. Let us
assume that $G(p)$ is a continuous function on $\R^N$.
Eq. (\ref{pde}) is equivalent to
\begin{equation}
L\frac{{\rm d}}{{\rm d}L}
\left\{ L^{-\sigma} F(Lp)\right\}=
L^{-\sigma} G(Lp).
\end{equation}
A special solution to (\ref{pde}) is then given by
\begin{equation}
F(p)=\int_0^1\frac{{\rm d}L}L L^{-\sigma}G(Lp).
\label{sol}
\end{equation}
Let us require the solution to be a continuous differentiable
function on $\R^N$. Then we notice the following facts:\\[2mm]
{\it I) There exists a unique solution to (\ref{pde}).
II) It is given by (\ref{sol}).}\\[2mm]
First, suppose that we have two different solutions $F_1(p)$
and $F_2(p)$ of (\ref{pde}) which are both continuous
differentiable. Their difference satisfies
\begin{equation}
\left\{p\frac{\partial}{\partial p}-\sigma\right\}
\left(F_1(p)-F_2(p)\right)=0.
\label{pro}
\end{equation}
The only solution to this equation, which is regular at the
origin in the case $\sigma <0$, is zero. Second, (\ref{sol})
is continuous differentiable and is a solution to (\ref{pde}).

Thus we can substitute initial or boundary data by regularity
to obtain a unique solution. Notice that its value at zero
is $-\sigma F(0)=G(0)$, as can be seen from both (\ref{pde})
and (\ref{sol}).

\subsubsection{Relevant Case}

The relevant case is defined by $\sigma >0$. In this case we
cannot use (\ref{sol}) because the integral diverges, unless
$G(p)$ provides a sufficiently high power of $p$. The trick is
to perform a Taylor expansion with remainder term to high
enough order. Let us assume that $G(p)$ is $(\sigma+1)$-times
continuous differentiable.

Let $\alpha =(\alpha_1,\ldots,\alpha_N)\in\N^N$ be an integer
valued multi-index. Define $|\alpha|=\alpha_1+\dots +\alpha_N$,
$\alpha !=\alpha_1!\dots\alpha_N!$, and $p^\alpha=p_1^{\alpha_1}
\dots p_N^{\alpha_N}$. Then $F(p)$ is a solution to (\ref{pde})
iff its derivatives with $\sigma\leq |\alpha|$ satisfy
\begin{equation}
\left\{p\frac{\partial}{\partial p}-(\sigma-|\alpha|)\right\}
\frac{\partial^{|\alpha|}F}{\partial p^\alpha}(p)=
\frac{\partial^{|\alpha|}G}{\partial p^\alpha}(p).
\label{tay}
\end{equation}
Thus each momentum derivative reduces the power counting
parameter by one unit. Solutions to (\ref{tay}) which are
regular at the origin satisfy
\begin{equation}
-(\sigma-|\alpha|)
\frac{\partial^{|\alpha|}F}{\partial p^\alpha}(0)=
\frac{\partial^{|\alpha|}G}{\partial p^\alpha}(0).
\label{coe}
\end{equation}
Their Taylor coefficients are all determined except for those
with $\sigma-|\alpha|=0$, the marginal ones. Let us assume
that
\begin{equation}
\frac{\partial^{|\alpha|}G}{\partial p^\alpha}(0)=0,\quad
\sigma-|\alpha|=0.
\end{equation}
Then (\ref{pde}) can be solved in terms of a Taylor
expansion of order $\sigma$ with remainder,
\begin{equation}
F(p)=
\sum_{|\alpha|\leq\sigma}
\frac{p^\alpha}{\alpha !}
\frac{\partial^{|\alpha|}F}{\partial p^\alpha}(0)+
\sum_{|\alpha|=\sigma+1}
\frac{p^\alpha}{\alpha !}
\int_0^1 {\rm d}t(1-t)^\sigma
\frac{\partial^{|\alpha|}F}{\partial p^\alpha}(tp).
\end{equation}
The derivatives of order $\sigma +1$ have negative
power-counting and are integrated as above. Let us require
the solution to be $(\sigma +1)$-times continuous differentiable.
Then we have: \\[2mm]
{\it I) There exists a set of solutions to (\ref{pde}) which
can be parametrized by its Taylor coefficients with $\sigma=
|\alpha|$.
II) The relevant Taylor coefficients with $\sigma<|\alpha|$
are uniquely determined by (\ref{coe}).
III) The Taylor remainder is reconstructed from the $(\sigma+1)$th
derivatives. It is unique and follows from (\ref{sol}).}
%

\subsection{Large Momentum Bound}

We prove a large momentum bound for the solution to the
renormalization group PDE under the assumption of a large
momentum bound on the inhomogeneous side. We choose an
$L_{\infty,\epsilon}$-norm for some $\epsilon >0$. It is
rather wasteful but suffices to prove finiteness of the
bilinear renormalization group bracket.

\subsubsection{Irrelevant Case}

Let $\sigma <0$. Suppose that the function $G(p)$ in
(\ref{pde}) has a finite $L_{\infty,\epsilon}$-norm
\begin{equation}
\| G\|_{\infty,\epsilon}=
\sup_{p\in\R ^N}
\left\{ |G(p)| e^{-\epsilon |p|} \right\}<\infty.
\label{normest}
\end{equation}
Then the solution (\ref{sol}) inherits an
$L_{\infty,\epsilon}$-bound. From
\begin{equation}
|F(p)| e^{-\epsilon |p|}\leq
\int_0^1 \frac{{\rm d}L}{L} L^{-\sigma}
|G(Lp)| e^{-\epsilon |p|}
\leq
\int_0^1 \frac{{\rm d}L}{L} L^{-\sigma}
e^{-(1-L)\epsilon |p|}
\| G\|_{\infty,\epsilon}
\end{equation}
it follows that
\begin{equation}
\| F\|_{\infty,\epsilon}\leq
\frac {1}{-\sigma}
\| G\|_{\infty,\epsilon}.
\label{bound}
\end{equation}
Eq. (\ref{bound}) shows that the irrelevant solution to the
renormalization group PDE is not only finite but also
decreases in the $L_{\infty,\epsilon}$-norm.

\subsubsection{Marginal Case}

Let $\sigma =0$. In this case we assemble $F(p)$ using a
first order Taylor formula. Suppose then that we have
$L_{\infty,\epsilon}$-bounds on the first derivatives
\begin{equation}
\| G_{\mu}\|_{\infty,\epsilon}=
\sup_{p\in\R ^N} \left\{
\left\vert \frac{\partial}{\partial p^{\mu}} G(p) \right\vert
e^{-\epsilon |p|} \right\} <\infty.
\label{derone}
\end{equation}
If $F(p)$ is marginal, then its first derivatives are
irrelevant with scaling dimension minus one. It follows that
\begin{equation}
\| F_{\mu}\|_{\infty,\epsilon} \leq
\| G_{\mu}\|_{\infty,\epsilon}.
\end{equation}
Therefrom it follows that
\begin{align}
|F(p)| e^{-\epsilon |p|} &\leq
|F(0)| e^{-\epsilon |p|}+
\sum_{\mu} |p_\mu |
\int_0^1 {\rm d}t
| F_\mu (tp)| e^{-\epsilon |p|}
\nonumber \\
&\leq
|F(0)|+
\sum_{\mu} |p_\mu |
\int_0^1 {\rm d}t
e^{-(1-t)\epsilon |p|}
\| F_\mu\|_{\infty,\epsilon}.
\end{align}
The result is an $L_{\infty,\epsilon}$-bound
\begin{equation}
\| F\|_{\infty,\epsilon}
\leq |F(0)|+
\frac {1}{\epsilon} \sum_{\mu}
\| F_\mu\|_{\infty,\epsilon}.
\end{equation}
This estimate is not uniform in $\epsilon$. It works for
$\epsilon$ arbitrary small, but the bound grows with an
inverse power of $\epsilon$. The large momentum growth
is a consequence of the split in derivatives and Taylor
remainder.

\subsubsection{Relevant Case}

Let $\sigma >0$. This case requires a generalization of
the bound in the marginal case. The Taylor expansion is
pushed to order $\sigma +1$. Then the derivatives become
irrelevant. We assume $L_{\infty,\epsilon}$-estimates on
all derivatives
\begin{equation}
\| G_\alpha \|_{\infty,\epsilon} =
\sup_{p\in\R ^N} \left\{
\left\vert \frac{\partial ^{|\alpha |}}
{\partial p^{\alpha}} G(p)\right\vert
e^{-\epsilon |p|}\right\} <\infty
\end{equation}
of order $|\alpha |=\sigma +1$. Since they are irrelevant
with scaling dimension minus one it follows that the
corresponding derivatives of $F(p)$ obey
\begin{equation}
\| F_\alpha \|_{\infty,\epsilon}\leq
\| G_\alpha \|_{\infty,\epsilon},
\end{equation}
and are also $L_{\infty,\epsilon}$-bounded. From the
Taylor formula it then follows that
\begin{align}
|F(p)| e^{-\epsilon |p|} &\leq
\sum_{|\alpha |\leq\sigma}
\frac {|p^\alpha |}{\alpha !} e^{-\epsilon |p|}
| F_\alpha (0)|+
\sum_{|\alpha|=\sigma +1}
\frac {|p^\alpha |}{\alpha !}
\int_0^1{\rm d}t (1-t)^\sigma
|F_\alpha (tp)| e^{-\epsilon |p|}
\nonumber \\ &\leq
\sum_{|\alpha |\leq\sigma}
\frac {|p|^{|\alpha |}}{\alpha !}
| F_\alpha (0)|+
\sum_{|\alpha|=\sigma +1}
\frac {|p|^{\sigma +1}}{\alpha !}
\int_0^1{\rm d}t (1-t)^\sigma
e^{-(1-t)\epsilon |p|}
\|F_\alpha \|_{\infty,\epsilon},
\end{align}
and thus
\begin{equation}
\| F \|_{\infty,\epsilon}\leq
\sum_{|\alpha |\leq \sigma} \frac {1}{\alpha !}
A_{\epsilon,|\alpha |} |F_{\alpha}(0)|+
\sum_{|\alpha |=\sigma+1} \frac {1}{\alpha !}
B_{\epsilon, \sigma+1} \| F_\alpha \|_{\infty,\epsilon},
\end{equation}
with constants
\begin{equation}
A_{\epsilon,|\alpha|}=\sup_{p\in\R ^N}
\left\{ |p|^{|\alpha |} e^{-\epsilon |p|}\right\},\quad
B_{\epsilon, \sigma +1}=\frac {\Gamma (\sigma +1)}
{\epsilon^{\sigma +1}}.
\end{equation}
Thus we again have an $L_{\infty,\epsilon}$-bound on
the function $F(p)$. This completes the large momentum bound
on $F(p)$. Exactly the same strategy applies to the derivatives
of $F(p)$ as well. The irrelevant derivatives inherit immediately
large momentum bounds. The relevant derivatives require Taylor
expansions. We omit to spell out explicitely the necessary
bounds on the derivatives of $G(p)$.

\subsection{Iteration and Regularity}

The iterative scheme determines order by order $\beta^{(s)}$,
$\zeta^{(s-1)}$, $\mu^{(s)}$, and the irrelevant remainders
$\wt{\VV}^{(s)}_{irr,2n}(p_1,\ldots,p_{2n-1})$. It is finite
to all orders of perturbation theory because of the following
iteration of regularity. Suppose that we have shown the
following to all orders $s\leq r-1$: \\[2mm]
{\it I) $\beta^{(s)}$,
$\zeta^{(s-1)}$, and $\mu^{(s)}$ are finite numbers. II)
$\wt{\VV}^{(s)}_{irr,2n}(p_1,\ldots,p_{2n-1})$ is a
smooth function on $\R\times\cdots\times\R$ for all
$1\leq n\leq s+1$, symmetric in the momenta, and
$O(D)$-invariant. III)
$\|\wt{\VV}^{(s)}_{irr,2n,\alpha}\|_{\infty,\epsilon}$
is finite for all $\epsilon >0$, $1\leq n\leq s+1$, and
$|\alpha |\geq 0$. Here $\alpha$ is a multi-index which
labels momentum derivatives.} \\[2mm]
Then the same statements hold at order $s=r$. Since they
are trivially fulfilled to order one they iterate to all
orders of perturbation theory.

To prove the iteration of regularity we once more inspect
each step of the iterative scheme. First, the irrelevant
remainders $\wt{\KK}^{(r)}_{irr,2n}(p_1,\ldots,p_{2n-1})$
are smooth functions on $\R\times\cdots\times\R$,
symmetric under permutations and $O(D)$-invariant. They
and all their momentum derivatives satisfy
$L_{\infty,\epsilon}$-bounds. They are composed of two
contributions. The first immediately inherits a bound from
the induction hypothesis. The second is a sum of
renormalization group brackets of lower orders.
Therefore, they consist of multiple convolutions with
propagators. The integrals converge, are smooth functions
of the external momenta, and satisfy
$L_{\infty,\epsilon}$-bounds. Second, we have linear
equations for the coefficients $\beta^{(r)}$,
$\zeta^{(r-1)}$, and $\mu^{(r)}$ with finite coefficients.
Third, the integration of the inhomogeneous renormalization
group PDEs, yields solutions with the desired properties.

\section{Conclusions}

The aim of perturbative renormalization theory is to derive
power series expansions for Green's functions which are free
of divergencies. The BPHZ theorem states that this can be
accomplished by writing the Green's functions in terms of
renormalized parameters. An elegant proof of the BPHZ theorem
was given by Callan \cite{C76}. A polished version of
which is due to Lesniewski \cite{L83}. Their method is
similar to ours in that it is based on renormalization group
equations for the renormalized Green's functions, the
Callan-Symanzik equations. In some sense (\ref{basic}) is
a Wilson-analogue of the Callan-Symanzik equations.
The method proposed here is different in that it does {\it not}
resort to any kind of graphical analysis, not to analysis
of sub-graphs, and not to skeleton expansions.

A new generation of proofs of the BPHZ theorem was
initiated with the work of Polchinski \cite{P84}. His proof
has been simplified further by Keller, Kopper, and
Salmhofer \cite{KKS90}. Their approach is similar to the
method advocated here in that it is based on Wilson's
exact renormalization differential equation. The details
are however quite different. The main difference is that
Polchinski begins with a cutoff theory. He then shows how
the cutoff can be removed in a way such that the effective
interaction remains finite. Our method directly 
addresses the limit theory without cutoffs, {\it expressed}
in terms of a renormalization group transformation with cutoffs.
In some sense we are here simultaneously changing Polchinski's
renormalization conditions and integrating an amount of
fluctuations. Unlike Polchinski and followers we use a
renormalization group differential equation with dilatation
term. A way to think of (\ref{basic}) is as a renormalization
group fixed point of a system which has been enhanced by one
degree of freedom, the running coupling. This fixed point
problem can only be formulated with rescaling and with
dilatation term.

Another renormalization group approach to renormalized
perturbation theory comes from Gallavotti \cite{G85,GN85}
and collaborators. Pedagocial accounts of tree expansions
can be found in \cite{BG95,FHRW88}.
There the result of renormalization is
expressed in terms of a renormalized tree expansion. The
program of \cite{Wi96} with an iterated transformation
with fixed $L$ is related to the tree expansion. Both are
built upon a cumulant expansion for the effective interaction.
The renormalization procedure is however quite different.
Like Polchinski, Gallavotti starts from a cutoff theory.
It is organized in terms of trees, which describe the
sub-structure of divergencies in Feynman diagrams. The
divergencies are transformed into a flow of the non-irrelevant
couplings. This part is similar to ours. The basic difference
with Gallavotti is that we do not organize our
expansion in terms of trees. A hybrid approach between
Polchinski and Gallavotti is due to Hurd \cite{H89}.

An important question is wether this construction of
renormalized trajectories extends beyond perturbation
theory.\footnote{It certainly works in the cases where
perturbation theory converges.} Another important question
is wether it extends to renormalized trajectories at
non-trivial fixed points. We hope to return with answers
to these questions in the future.



\begin{thebibliography}{XXXXX}
%
\bibitem[BG95]{BG95} G.\,Benfatto, G.\,Gallavotti,
Renormalization group, Physics Notes No. 1, Princeton
University Press 1995
%
\bibitem[C76]{C76} C.\,G.\,Callan, Introduction to
renormalization theory, Les Houches Lecture Notes
1975, 41-77, R.\,Balian and J.\,Zinn-Justin eds.
%
\bibitem[FHRW88]{FHRW88} J.\,S.\,Feldman, T.\,R.\,Hurd,
L.\,Rosen, J.\,D.\,Wright, QED: A proof of renormalizability,
Lecture Notes in Physics 312, Springer Verlag 1988
%
\bibitem[G85]{G85} G.\,Gallavotti, Renormalization theory
and ultraviolet stability for scalar fields via renormalization
group methods, Rev. Mod. Phys. Vol. 57 No. 2 (1985) 471-562
%
\bibitem[GN85]{GN85} G.\,Gallavotti and F.\,Nicolo,
Renormalization in four dimensional scalar fields I,
Commun. Math. Phys. 100 (1985) 545-590; Renormalization
Renormalization in four dimensional scalar fields II,
Commun. Math. Phys. 101 (1985) 247-282
%
\bibitem[GJ87]{GJ87} J.\,Glimm and A.\,Jaffe, Quantum Physics,
Springer Verlag 1987
%
\bibitem[H89]{H89} T.\,Hurd, A renormalization group proof
of perturbative renormalizability, Commun. Math. Phys. 124
(1989) 153-168
%
\bibitem[KKS90]{KKS90} G.\,Keller, C.\,Kopper, M.\,Salmhofer,
Perturbative renormalization and effective Lagrangeans,
MPI-PAE/PTH 65/90
%
\bibitem[L83]{L83} A.\,Lesniewski, On Callan's proof of the
BPHZ theorem, Helv. Phys. Acta, Vol. 56 (1983) 1158-1167
%
\bibitem[P84]{P84} J.\,Polchinski, Renormalization and
effective Lagrangeans, Nucl. Phys. B231 (1984) 269-295
%
\bibitem[W71]{W71} K.\,Wilson, Renormalization group and
critical phenomena I and II, Phys. Rev. B4 (1971) 3174-3205
%
\bibitem[We76]{We76} F.\,J.\,Wegner, The critical state,
general aspects, in Phase Transitions and Critical Phenomena
Vol. 6, C.\,Domb and M.\,S.\,Green eds., Academic Press 1976
%
\bibitem[Wi88]{Wi88} C.\,Wieczerkowski, Symanzik's improved
actions from the viewpoint of the renormalization group,
Commun. Math. Phys. 120, 149-176 (1988)
%
\bibitem[Wi96]{Wi96} C.\,Wieczerkowski, The renormalized
$\phi^4_4$-trajectory by perturbation theory in the running
coupling, hep-th/9601142
%
\bibitem[WK74]{WK74} K.\,Wilson and J.\,Kogut,
The renormalization group and the $\epsilon$-expansion,
Phys. Rep. C12 No. 2 (1974) 75-200
%
\end{thebibliography}
\end{document}